\documentclass[prl,aps,reprint,amsmath,amssymb,longbibliography, superscriptaddress]{revtex4-2}

\usepackage{ajtex-revtex}

\def\nem{\mathrm{nem}}
\def\tot{\text{tot}}
\def\diss{\text{nhs}}
\newcommand\env{{\mathsmaller{\mathsf E}}}
\newcommand\fuel{{\mathsmaller{\mathsf F}}}

\begin{document} 

%suggestion
\title{On the temperature of an active nematic}
%Energy fluctuations and long range correlations in active nematics

\author{Jay Armas}\email{j.armas@uva.nl}

\affiliation{Institute for Theoretical Physics, University of Amsterdam, 1090 GL Amsterdam, The Netherlands}
\affiliation{Dutch Institute for Emergent Phenomena, 1090 GL Amsterdam, The Netherlands}
\affiliation{Institute for Advanced Study, University of Amsterdam, Oude Turfmarkt 147, 1012 GC Amsterdam, The Netherlands}
\affiliation{Niels Bohr International Academy, The Niels Bohr Institute, University of Copenhagen,
Blegdamsvej 17, DK-2100 Copenhagen \O{}, Denmark}

\author{Akash Jain}\email{a.jain2@uva.nl}
\author{Ruben Lier}\email{r.lier@uva.nl}

\affiliation{Institute for Theoretical Physics, University of Amsterdam, 1090 GL Amsterdam, The Netherlands}
\affiliation{Dutch Institute for Emergent Phenomena, 1090 GL Amsterdam, The Netherlands}
\affiliation{Institute for Advanced Study, University of Amsterdam, Oude Turfmarkt 147, 1012 GC Amsterdam, The Netherlands}

\date{\today}

\begin{abstract}
We employ a novel hydrodynamic framework for active matter coupled to an environment to study the local temperature of an active nematic, assuming proximity to thermal equilibrium. We show that, due to the mechanosensitivity of fuel consumption, linearized temperature correlations in a homogeneous active nematic steady state remain unaffected by activity. However, we demonstrate that local shearing and twisting cause a confined active nematic undergoing a spontaneous flow transition to develop a distinctive inhomogeneous temperature profile, serving as a thermal signature of activity.
%
%%% Akash's Previous Version
% For active matter close to thermal equilibrium, Onsager’s relations require the fuel consumption to become mechanosensitive, influencing the relationship between activity and local temperature. Using a new hydrodynamic theory for thermal active nematics, we show that mechanosensitivity in energy balance prevents activity from affecting long-range temperature correlations in a homogeneous steady state. However, due to local shearing and twisting, we find that a confined active nematic undergoing a spontaneous flow transition develops an inhomogeneous temperature profile that signals activity.
%
%%% Ruben's Version
% For active matter close to thermal equilibrium, the Onsager relations dictate that fuel consumption becomes mechanosensitive in the presence of activity, which has consequences for the relation between activity and temperature. In this work, we use a new framework of hydrodynamics for thermal active matter to show that the active nematic contributions to energy balance cancel out at linear order, preventing active nematicity from affecting thermal corrections. However, we show that at second order in perturbations, an active nematic in a confined geometry undergoing a spontaneous flow transition displays a temperature profile that signals activity.  
\end{abstract} 

\maketitle

Active matter consists of microscopic, fuel-consuming agents, enabling it to exhibit macroscopic behavior that is distinct from ordinary passive matter. In particular, when active matter admits a continuum hydrodynamic description for collective transport, it possesses the remarkable ability to circumvent various constraints that govern passive systems, arising from the local second law of thermodynamics, Onsager’s reciprocity relations, and fluctuation-dissipation theorems (FDTs).

The transport phenomena in active matter become particularly striking when some form of spontaneous order is present, such as in nematic liquid crystals, which exhibit spontaneous rotational order~\cite{Chaikin_Lubensky_1995,de1993physics}. \emph{Active nematics} constitute one of the most widely studied classes of active matter, from both experimental and theoretical perspectives~\cite{Doostmohammadi2018,Saw2017,Assante2023,RevModPhys.85.1143,Van_Saarloos2024}. In particular, they exhibit distinctive turbulent states that do not arise in ordinary passive nematics~\cite{Sanchez2012, PhysRevLett.110.228101, Alert_2020, turbulencealert}.
Active nematics are realized in a variety of physical systems, such as bacterial populations~\cite{PhysRevLett.108.098102, PhysRevE.94.050602, PhysRevE.95.020601, PhysRevX.7.011029}, microtubule-motor protein mixtures~\cite{Ndlec1997,Butt2010}, swarming sperm cells~\cite{PhysRevE.92.032722}, and epithelial cells~\cite{Epithelial1, Epithelial2, Epithelial3}. Hydrodynamics provides a promising theoretical framework for describing active nematics, wherein the hydrodynamic theory of ordinary passive nematics~\cite{beris1994thermodynamics} is extended to incorporate the effects of fuel consumption \cite{PhysRevLett.89.058101, PhysRevLett.92.078101, annurev:/content/journals/10.1146/annurev.fl.24.010192.001525}. 

Hydrodynamics describes the near-equilibrium macroscopic behavior of a varied classes of physical systems. It governs the evolution of collective variables, such as energy, momentum, and mass, as well as Goldstone modes of spontaneously broken symmetries~\cite{1972PhRvA...6.2401M}.
A central assumption of hydrodynamics is \emph{local thermal equilibrium}, which breaks down in the presence of microscopic active agents that drive the system away from equilibrium. However, when active effects remain reasonably small and comparable to the local gradients away from thermal equilibrium, activity can be introduced perturbatively within the hydrodynamic framework~\cite{grootmazur,J_licher_2018,PhysRevE.60.2127}. 

Extending the framework of hydrodynamics to active matter requires accounting for the fuel-consuming processes inherent to active agents, which collectively serve as a local energy source supporting activity~\cite{J_licher_2018}. As such, fuel consumption leads to a continuous build-up of energy in the system, preventing it from reaching a stable steady state. Consequently, such a framework inevitably cannot make long-time predictions for temperature. To address this issue, one must recognize that active matter is not only defined by the presence of fuel-consuming agents but also by its ability to dissipate excess energy into the environment.

In earlier work, we incorporated this idea into the hydrodynamics of active matter by allowing the system to exchange of energy and entropy with an environment at a fixed temperature $T_\env$~\cite{Armas:2024iuy}. This restores the balance of energy, thereby allowing for non-equilibrium steady states maintained at a temperature above $T_\env$. The loss of entropy to the environment also modifies the local second law of thermodynamics, making it possible for active matter to acquire unique transport properties that are not present in passive systems~\cite{Armas:2024iuy}.

The conversion of chemical energy to kinetic energy sources the energy balance equation of the fluid, and Onsager reciprocity dictates that his sourcing of energy is \emph{mechanosensitive}, i.e. sensitive to the state of the system~\cite{J_licher_2018, JULICHER20073, PhysRevLett.92.078101, Callan-Jones_2011}. We show that, as a consequence, linear temperature correlations in a homogeneous active nematic steady state are entirely agnostic of activity, which is unlike the density correlator in dry active matter \cite{Basu2008,SRamaswamy_2003,PhysRevE.97.012707}. Nonetheless, temperature dynamics exhibit non-trivial active signatures when the nematic enters an inhomogeneous state induced by activity. To study this, we compute the local temperature profile of an active nematic in a confined geometry close to the spontaneous flow transition~\cite{Voituriez_2005}. We unravel an interesting qualitative interplay between the local temperature profile and the length scale of energy relaxation to the environment\footnote{A work that appeared recently computes the thermal profile for pipe flows in the presence of an imposed activity gradient which manifests as a pressure gradient \cite{10.1063/5.0258996}. The key difference with our approach to temperature is that the energy balance equation we consider accounts for the fuel consumption that gives rise to activity. Additionally, we consider a different type of active matter and assume that it is only activity that drives the system out of equilibrium.}. Our work provides compelling evidence of how the new hydrodynamic framework can enable us to understand the thermal characteristics of active matter. 

% Using this novel hydrodynamic framework, we study linear temperature fluctuations in a homogeneous active nematic steady state. 

\vspace{0.5em}
\noindent
\textit{Passive hydrodynamics---}%
To set the stage, let us begin with a passive two-dimensional incompressible fluid~\cite{landau1959fluid}. The hydrodynamic variables consist of the conserved energy density $\epsilon$ and momentum density $\pi^i$, with $i,j,\ldots \in \{x,y\}$ denoting spatial indices. Correspondingly, we have the conservation equations
\begin{align}\label{eq:energybalance}
    \dow_t{\epsilon} +  \partial_i\epsilon^i  
    &=   0 ~, \qquad
    \dow_t \pi^i -  \dow_j t^{ij}
    =  0 ~,
    \end{align}
where $\epsilon^i$ and $t^{ij}$ denote the energy flux and stress tensor respectively. Incompressibility requires $\partial_i v^i =0$, where $v^i=\pi^i/\rho_0$ is the fluid velocity and $\rho_0$ is the constant mass density. Identifying the comoving energy density $\varepsilon = \epsilon-\half \rho_0 v^2$ and flux $\varepsilon^i=\epsilon^i + \half\rho v^2 v^i +  t^{ji}v_j$, the constitutive equations for the passive fluid are given by
\begin{align}\label{eq:fluxes}
    \varepsilon^i
    &= \varepsilon v^i + \varepsilon^i_\diss~, \qquad
    t^{ij}
    = - \pi^i v^j - p\,\delta^{ij} + t^{ij}_\diss~.
\end{align}
Here $p$ and $T$ are the local fluid pressure and temperature, while $\varepsilon^i_\diss$ and $t^{ij}_\diss$ denote the dissipative fluxes.

To fix the dissipative constitutive relations, let us consider the dissipation rate of the fluid $\Theta \equiv \dow_t s + \dow_i s^i$, where $s$ is the entropy density and $Ts^i=Tsv^i + \varepsilon^i_\diss$ is the heat flux. Using thermodynamic identities and conservation equations, we find 
\begin{align}\label{eq:Theta-passive}
    T\Theta = 
    - \frac{1}{T} \varepsilon^i_\diss\dow_iT
    + t^{ij}_\diss u_{ij}~,
\end{align}
where $u_{ij} =\frac{1}{2} ( \partial_{i} v_{j } + \partial_{j} v_{i })$ is the fluid shear tensor.
Since the dissipation rate should be locally non-negative, the nonhydrostatic fluxes must take the form
\begin{align}\label{eq:dissipative-passive}
    \varepsilon^i_\diss
    = - \kappa\, \partial^i T + \xi^i_{\epsilon}~, \qquad 
    t^{ij}_\diss
    = 2\eta\, u^{ij} + \xi^{ij} _{\pi}~.
\end{align}
where $\kappa$ and $\eta$ are the thermal conductivity and viscosity of the fluid respectively. Requiring $\Theta\geq 0$ leads to the second-law constraints $\kappa,\eta\geq 0$. 

We have also included the stochastic fluxes $\xi$'s in \cref{eq:dissipative-passive} to account for thermal fluctuations \cite{1987xi,Fox1970,PhysRev.91.1512}.
These are defined to have zero mean, $\langle \xi(x,t)\rangle =0$, but nonzero short-ranged variance $\langle \xi(x,t)\xi(x',t')\rangle \propto \delta^{(2)}(x-x')\delta(t-t')$. In local thermal equilibrium, the magnitude of the variance is fixed by FDT in terms of the dissipative transport coefficients $\kappa$ and $\eta$; see \cref{app:noisevariance} for a review.

We can use this hydrodynamic model to derive the temperature dynamics on a passive incompressible fluid. Ignoring stochastic noise, we have \cite{chandrasekhar1981hydrodynamic}
\begin{equation}\label{eq:temperature-passive}
    c_v\frac{\df}{\df t} T
    - \partial_i\big(\kappa\,\dow^iT\big)
    = 2\eta\, (u_{ij})^2,
\end{equation}
where $\df/\df t = \dow_t + v^i\dow_i$ is the material derivative along the fluid and $c_v = \partial \epsilon/\partial T$ is the heat capacity. At linearized order, temperature fluctuations decouple and exhibit diffusional dispersion relations $\omega = -iD_\epsilon k^2$, were $D_{\epsilon} = \kappa/c_v$ is the energy diffusion coefficient and $\omega,k^i$ are the frequency and wavevector of linearized perturbations. More generally, however, inhomogeneous shearing of the fluid velocity affects temperature via non-linear interactions.

% We have defined , with 

\vspace{0.5em}
\noindent
\textit{Hydrodynamics with fuel consumption---}%
We now introduce fuel consumption into our hydrodynamic model, which provides the energy source needed for supporting activity~\cite{J_licher_2018}. As we formalized in~\cite{Armas:2024iuy}, for such models to admit an active steady state maintained at a fixed temperature, we need to also allow for an outflow of energy to the environment. In the present case, the environment may model a fixed substrate on which the two-dimensional active nematic lies.
These new ingredients are summarized in the energy balance equation~\footnote{The energy balance equation can also be recast in terms of total energy, including the fluid and fuel contributions, which is conserved when the outflow of energy to the environment is absent~\cite{J_licher_2018}. However, the non-conservative coupling to the environment is crucial to maintain active steady states; see appendix A of~\cite{Armas:2024iuy} for more details.}
\begin{align}  \label{eq:energybalance12}
    \dow_t{\epsilon} +  \partial_i\epsilon^i  
    &= r_\fuel \Delta \mu
    - r_\env  \kb  T_\env~,
\end{align}
where we have introduced the fuel reaction rate $r_\fuel$ and heat loss rate $r_\env$, together with the corresponding chemical potential difference $\Delta \mu$ and environment temperature $T_{\env}$. Throughout this work, we will assume $\Delta \mu$ and $T_{\env}$ to be constant and positive.

We can again employ the second law of thermodynamics to determine the form of the rates $r_{\fuel,\env}$. In the presence of energy outflow to the environment, the dissipation rate $\Theta$ is no longer necessarily non-negative. However, the total dissipation rate $\Theta_{\tot}=\Theta+\kb r_\env$, which also includes entropy loss to the environment,
% , i.e.
% \begin{align} \label{eq:dissipationrate}
%       \Theta_{\tot} \equiv \dow_t s + \dow_i s^i 
%       + \kb r_\env ~,     
% \end{align}
should still be non-negative \cite{Armas:2024iuy}. We find
\begin{align}\label{eq:second-law-diff111}
    T \Theta_\tot
    &= r_{\fuel} \Delta \mu  + r_{\env} \kb \Delta T
    - \frac{1}{T} \varepsilon^i_\diss\dow_iT
    + t^{ij}_\diss u_{ij} ~,
\end{align}
which leads to the constitutive relations for the rates
\begin{align}\label{eq:fuelterms}
    r_{\fuel}   
    &= \gamma_{\fuel } \Delta \mu 
    + \xi_{\fuel}~, \qquad
    r_{\env}   
    = \gamma_{\env } \kb \Delta T   
    + \xi_{\env}~.
\end{align}
Here $\Delta T=T-T_\env$ is the temperature differential between the system and the environment, $\gamma_{\fuel,\env}$ are some coefficients that control the respective rates, while $\xi_{\fuel,\env}$ represent the associated stochastic noise.
Requiring $\Theta_\tot\geq 0$ enforces the constraints $\gamma_{\fuel},\gamma_\env\geq0$, in addition to $\kappa,\eta\geq 0$.

Plugging the rates \eqref{eq:fuelterms} back into the energy balance equation \eqref{eq:energybalance12}, we can derive the temperature dynamics
\begin{equation}\label{eq:T-dynamics-fuel}
    c_v \frac{\df}{\df t} T 
    - \partial_i\big(\kappa\,\dow^iT\big)
    + c_v\Gamma_\epsilon \delta T
    = 2\eta\, (u_{ij})^2~,
\end{equation}
where $\Gamma_\epsilon=\kb^2T_\env\gamma_\env/c_v$ and $\delta T = T-T_0$, with
\begin{equation}\label{eq:eqb-temp}
    T_0
    = T_\env
    \lb 1 + \frac{\gamma_\fuel}{\gamma_\env}\frac{\Delta \mu^2}{\kb^2 T_\env^2} \rb~.
\end{equation}
It follows that the fluid only admits a steady state solution at temperature $T_0$, which is strictly above the environment temperature $T_\env$ in the presence of fuel consumption, while being equal to $T_\env$ when the fuel is turned off. Such a steady state solution does not exist when coupling to environment is absent, i.e. when $\gamma_\env\to0$, and leads to a steady growth of energy and temperature~\cite{J_licher_2018}. Since energy is not conserved, linearized temperature fluctuations near the steady state relax at characteristic time-scale $2\pi/\Gamma_\epsilon$, with dispersion relations $\omega = -i\Gamma_\epsilon - i D_\epsilon k^2$. We can also identify the characteristic length-scale $L_\epsilon = 2\pi\sqrt{D_\epsilon/\Gamma_\epsilon}$ at which the temperature fluctuations decay.

FDT only applies to systems fluctuating around thermal equilibrium and, as such, cannot be used in a non-equilibrium steady state. Nonetheless, we can tune the system close to thermal equilibrium by turning off the fuel source and then use FDT to calibrate the noise variances; see~\cite{Basu2008,sengersortiz}. We find (see \cref{app:noisevariance})
\begin{align}\label{eq:noisevariances13}
\Big\langle \xi_{\fuel}(x,t)\,\xi_{\fuel}(x',t')\Big\rangle   
&=  2\kb T_0 \gamma_{\fuel}  \delta^{(2)}(x-x')\delta (t - t' )~, \nn\\ 
\Big\langle  \xi_{\env}(x,t) \,   \xi_{\env}(x',t') \Big\rangle  &  =  2  \kb  T_0      \gamma_{\env } \delta^{(2)} (x - x' )  \delta (t - t' )~.
 \end{align}

\vspace{0.5em}
\noindent
\textit{Active nematics---}%
The effect of fuel consumption has been somewhat trivial in our hydrodynamic model thus far. The temperature evolution in \cref{eq:T-dynamics-fuel} is qualitatively only affected by energy outflow to the environment, which determines the relaxation rate $\Gamma_\epsilon$, while fuel consumption only affects the value of the steady state temperature $T_0$. By contrast, velocity evolution is not affected by fuel consumption at all. To probe more non-trivial signatures of activity, let us now generalize our hydrodynamic model to active nematics. We consider the symmetric-traceless nematic order parameter $Q_{ij}=S(n_in_j-\half\delta_{ij})$, 
% \begin{align}\label{eq:nematic-state}
%     Q_{ij }  =   \frac{S}{2} \begin{pmatrix}
%      \cos(2 \theta   ) &   \sin(2 \theta   )   \\ 
%     \sin(2 \theta  ) &  - \cos(2 \theta ) 
%     \end{pmatrix}  ~~ , 
% \end{align}
where the unit vector $n_i=(\cos\theta,\sin\theta)$ represents the orientation of nematic order and the scalar $S$ determines its strength. The nematic free energy density is given by
\begin{align}\label{eq:nematic-F}
    \mathcal{F}_{\text{nem}}  
    = \frac{K}{2} (\partial_k Q_{ij} )^2 
    +  \frac{a}{2}(Q_{ij})^2 
    + \frac{b}{2} (Q_{ij})^4~,
\end{align}
where $K$ is the Frank distortion coefficient. In general, the Frank coefficients for bend and splay distortions may differ~\cite{de1993physics}, but we focus on the isotropic case for simplicity. The parameters $a$, $b$ determine the state of the system. For $a>0$, the system exists in the isotropic state, $S=0$, without nematic order. Whereas, for $a<0$, the system transitions to the nematic state with $S=\sqrt{-a/b}$.

The nematic free energy \eqref{eq:nematic-F} contributes to the constitutive relations for energy flux and stress tensor as
\begin{align}\label{eq:activet}
    \varepsilon^i
    &=  
    \varepsilon v^i
    - K\dow^i Q^{kl} \frac{\df}{\df t} Q_{kl}
    + \varepsilon^i_\diss
    ~~, \nn\\
    t^{ij}
    &= 
    - \pi^i v^j +
    \big(\mathcal{F}_{\text{nem}} - p\big) \delta^{ij} 
    -  K\dow^i Q^{kl} \dow^j Q_{kl}  \nn\\
    &\qquad
     + Q^{i}{}_{k}  \mathcal{H}^{ jk  }
     - Q^{j}{}_{k}  \mathcal{H}^{ ik  }
     + t^{ij}_\diss~,
\end{align}
where $\cH^{ij}=-\delta\cF_{\text{nem}}/\delta Q_{ij}$. The equation of motion for $Q_{ij}$ can be expressed as
\begin{align} \label{eq:Qequation}
    \frac{\df}{\df t} Q_{ij} 
    + \Big(\Omega_{i}{}^k Q_{kj} - Q_{ik} \Omega^k{}_{j} \Big)
    =  \cV^{\diss}_{ij}~,  
\end{align}
where $\Omega_{ij} = \frac{1}{2} 
(\partial_{i} v_{j} - \partial_{j} v_{i}) $ is the fluid vorticity tensor that enters the co-rotational term in the parenthesis, while $\cV_{ij}^\diss$ denotes the dissipative nematic dynamics.

In the presence of active nematicity, together with the fuel source and energy sink, the total dissipation rate modifies from \cref{eq:second-law-diff111} to (see~\cref{app:noisevariance})
\begin{align}
\label{eq:second-law-diff1113}
\begin{split}
  T  \Theta_\tot  
  &= r_{\fuel} \Delta \mu  + r_{\env} \kb \Delta T \\
  &\qquad 
  - \frac{1}{T} \varepsilon^i_\diss \partial_iT
    + \tau^{ij}_\diss u_{ij}
    + \cV^\diss_{ij} \mathcal{H}^{ij}~. 
\end{split}
\end{align}
This leads to the dissipative constitutive relations \cite{J_licher_2018}
\begin{align}\label{eq:activet2}
    \varepsilon^i_\diss
    &=  
    -\kappa\dow^i T + \xi^i_\epsilon~~, \nn\\
    t^{ij}_\diss
    &= 
    2\eta\, u^{ij}
    - S \lambda'\mathcal{H}^{ij}
    - \zeta   Q^{ij}
    + \xi^{ij} _{\pi}~~, \nn\\
    \cV^{ij}_{\diss}
    &= \frac{1}{\sigma_Q}\cH^{ij}
    + S \lambda u^{ij}
    + \xi_Q^{ij}~~, \nn\\
    r_\fuel 
    &=  \gamma_\fuel   \Delta \mu 
 + \alpha_{\fuel }    Q^{ij}
    u_{ij }  +   \xi_{\fuel}~~,
\end{align}
where $\sigma_Q$ is the rotational viscosity,  
while $\lambda$ and $\lambda^{\prime}$ correspond to the flow-alignment parameter \cite{PhysRevA.46.4966,Giomi_2012}. In particular, the stress tensor contains an active term proportional to $Q^{ij}$, controlled by the coefficient $\zeta$, which is absent for passive nematics. We have also included a mechanosensitive coefficient $  \alpha_{\fuel }    $ in the fuel rate, whose relevance will be clear momentarily. The variances of respective stochastic fluxes is provided in \cref{app:noisevariance}.

Assuming that the microscopic system underlying the hydrodynamic model, including the fuel and environment components, is time-reversal invariant, Onsager's reciprocity relations require that $\lambda'=\lambda$, together with the mechanosensitivity constraint \footnote{In theory, one can also account for the active contribution $ \zeta $ in an Onsager reciprocal way by adding a mechanosensitive contribution to $r_{\env}$. This possibility was considered in \cite{Armas:2024iuy}.}
\begin{align}\label{eq:onsgaer-mechanosensitivity}
     \zeta   =   \alpha_{\fuel }    \Delta\mu~~.
\end{align}
In other words, the fuel burning rate $r_{\fuel}$ becomes mechanosensitive in the presence of activity~\cite{J_licher_2018}. Further imposing $\Theta_\tot \geq 0 $ leads to the constraint $\sigma_Q\geq 0$.

\vspace{0.5em}
\noindent
\textit{Active temperature dynamics---}%
Due to the mechanosensitivity of fuel consumption, temperature dynamics of an active nematic is not directly sensitive to activity. To wit, the energy balance equation leads to
\begin{equation}\label{eq:energybalance123}
    c_v \frac{\df}{\df t} T 
    - \partial_i\big(\kappa\,\dow^iT\big)
    + c_v\Gamma_\epsilon\delta T
    =
    2\eta\, (u_{ij})^2
     + \frac{1}{\sigma_Q }  (\cH_{ij})^2~,
\end{equation}
which is not affected by the active coefficient $ \zeta  $.
Had we not imposed the Onsager mechanosensitivity constraint \eqref{eq:onsgaer-mechanosensitivity}, the right-hand side would contain an additional term $\big(  \alpha_{\fuel } \Delta\mu   - \zeta   \big) Q^{ij} u_{ij}$, which would indeed be sensitive to $ \zeta  $. Because of this, temperature correlations of an active nematic linearized around a homogeneous steady state do not signal activity.

To see this, let us consider the steady state given by fixed values of $S=S_0\equiv \sqrt{-a/b}$ and $T=T_0$ given in \cref{eq:eqb-temp}, and arbitrary constant values of $\theta=\theta_0$ and $v^i=v^i_0$. Small wavevector perturbations of this state, with $k < k_c(\Delta\theta)$, are dynamically unstable in the presence of activity and drive the system into a state of nematic turbulence~\cite{PhysRevLett.89.058101}. Here 
\begin{align}\label{eq:stability-bound}
    k_c(\Delta\theta)^2
    = \frac{ \zeta   S_0}{D_Q}
    \frac{\big( 1 +\lambda\cos(2\Delta\theta)\big) \cos(2\Delta\theta)}{2\eta + S_0^2\sigma_Q 
    \big( 1 + \lambda^2 + 2\lambda\cos(2\Delta\theta)\big)},
\end{align}
and $\Delta\theta = \theta_0 - \arctan(k_y/k_x)$ is the relative angle between the wavevector and the nematic order and $D_Q=K/\sigma_Q$ is the nematic diffusion coefficient. See~\cref{sec:linear-instability} for a review. Nonetheless, it is possible to stabilize the homogeneous steady state, either by introducing momentum friction to a substrate \cite{PhysRevE.90.062307} or by accounting for finite system size effects~\cite{Giomi_2012, giomiroyal, Ramaswamy_2007}. In these scenarios, one may compute the correlation function of temperature in the homogeneous steady state, which may be observed with small-angle light scattering experiments \cite{sengersortiz,LI1994399}. Using the distribution of noise fields in \cref{app:noisevariance}, we find
\begin{align}
    &\Big\langle T (x,t) T (x',t')\Big\rangle
    = \frac{2\kb T_0^2}{c_v}\int\frac{\df\omega\df^2k}{(2\pi)^3}{\rm e}^{-i\omega(t-t')+ik\cdot(x-x')} \nn\\
    &\hspace{12em}
    \frac{D_{\epsilon}k^2+\Gamma_{\epsilon}}
    {\omega ^2+(D_{\epsilon}k^2+\Gamma_{\epsilon})^2}~,
\end{align}
which, as predicted, is purely relaxational and agnostic of the active-nematic coefficient $\zeta$.

\begin{figure}
    \centering    \includegraphics[width=1\linewidth]{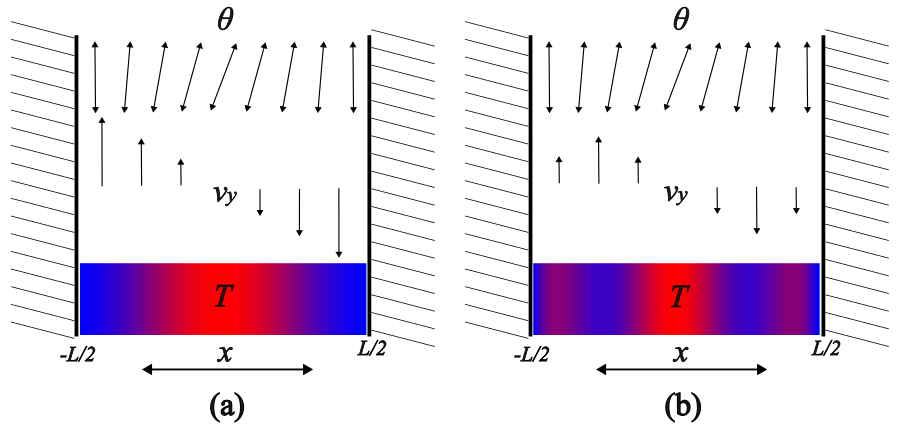}
    \caption{Schematic diagram of slab geometry depicting the nematic angle $\theta$, velocity $v_y$, and temperature $T$ along the $x$-axis. $\theta$ and $T$ are subject to Dirichlet boundary conditions, whereas $v_y$ is constrained by free slip and no-slip boundary conditions. We have taken $L_{\epsilon} = 0.4 L $.}
    \label{fig:enter-label}
\end{figure}

\vspace{0.5em}
\noindent
\emph{Temperature near spontaneous flow transition---}%
Activity may indirectly heat up the fluid via shearing and twisting terms on the right-hand side of \cref{eq:energybalance123}.
This occurs at the onset of nematic turbulence, when activity destabilizes and destroys the homogeneous steady state. To demonstrate this, we consider the well-known spontaneous flow transition that arises for active nematics in a confined geometry~\cite{Voituriez_2005}. Consider a contractile ($ \zeta < 0$) active nematic confined in a two-dimensional slab geometry, $x\in(-L/2,L/2)$ and $y\in(-\infty,\infty)$, where the nematic orientation is anchored at the boundaries, $\theta(\pm L/2) = \pi/2 $ (see \cref{fig:enter-label}). A similar setup holds for extensile active nematics ($ \zeta > 0$) with $\theta (\pm L/2) = 0$~\cite{2024arXiv240316841L}. 

We wish to look for steady state solutions to this problem. There is the trivial homogeneous steady state solution $S=S_0$, $T=T_0$, $v_y=0$, and $\theta=\pi/2$. As it turns out, with free-slip (\textsf{FS}) boundary conditions, $t_{xy} (\pm L/2) = 0$, the homogeneous state is dynamically stable for slabs narrower than the critical width, $L<L_c$, where $L_c = \pi/k_c(\pi/2)$~\cite{Voituriez_2005}. This is because finite slab width enforces a minimum wavevector in the $x$-direction, $k_x^{\text{min}} =\pi/L$, which is already above the stability bound in \cref{eq:stability-bound}. As we cross to $L>L_c$, the $k_x^{\text{min}}$ mode becomes unstable and the system undergoes a spontaneous flow transition to an inhomogeneous state \footnote{This is assuming that the transition is continuous. For certain values of $\lambda$, the transition can become subcritical and discontinuous \cite{lavi2024nonlinearspontaneousflowinstability}.}. Similarly, for no-slip (\textsf{NS}) boundary conditions, $v_y(\pm L/2)=0$, the homogeneous state is stable for $L<2L_c$~\cite{Voituriez_2005}.

Sufficiently close to criticality, we can perturbatively construct the stable inhomogeneous solution~\cite{Voituriez_2005}. See \cref{app:confinedgeometry} for a review. In particular, the leading perturbative shear profile is given as
\begin{equation}\label{eq:shear-profile}
    u_{xy}^{\textsf{FS}}
    \approx
    A_u\cos(\pi x/L)~, \quad
    u_{xy}^{\textsf{NS}}
    \approx 
    2A_u\cos(2\pi x/L)~,
\end{equation}
where the amplitude $A_u$ is proportional to $\sqrt{L-L_c}$ (for \textsf{FS}) and $\sqrt{L-2L_c}$ (for \textsf{NS}), which is sensitive to the non-linear structure of the theory~\cite{Voituriez_2005}. The full non-linear solution farther from the critical point can be constructed numerically~\cite{Giomi_2012, Edwards_2009, PhysRevE.76.031921}. 

The energy balance equation \eqref{eq:energybalance123} implies that the temperature across the slab is sourced by the square of the shear profile; at leading perturbative order
\begin{equation}\label{eq:non-linear-heating}
    \Big( \Gamma_\epsilon - D_\epsilon \partial_x^2 \Big) \delta T
    \approx - \frac{4\tilde\eta}{c_v} u_{xy}^2~,
\end{equation}
where $\tilde\eta = \eta + \half S_0^2\sigma_Q(1-\lambda)^2$ is the effective viscosity, modified due to nematic distortions. Solving \cref{eq:non-linear-heating} with Dirichlet boundary conditions, $T(\pm L/2)=T_0$, yields
\begin{align}\label{eq:T-profile}
    \delta T_{\textsf{FS}}
    &\approx \frac{L_\epsilon^2 A_T}{1+L_\epsilon^2/L^2}
    \left(
    \frac{L_\epsilon^2}{2L^2}f(x)
    + \cos^2(\pi x/L)
    \right)~, \nn\\
    \delta T_{\textsf{NS}}
    &\approx \frac{4L_\epsilon^2 A_T}{1+4L_\epsilon^2/L^2}
    \left(\frac{L^2+2L_\epsilon^2}{L^2} f(x)
    - \sin^2(2\pi x/L) \right),
\end{align}
where $f(x)=1-\cosh(2\pi x/L_\epsilon)\sech(\pi L/L_\epsilon)$ and we have defined $A_T = A_u^2\tilde\eta/(\pi^2\kappa)$ (see \cref{app:temperature-profile}). The temperature profile with Neumann boundary conditions, $\dow_x T(\pm L/2)=0$, is given in \cref{app:temperature-profile}. In \cref{app:estimate}, we provide a heuristic estimate of the magnitude of temperature modulations for the experimental setup of \cite{duclos2018spontaneous}.

\begin{figure}[t]
    \centering 
    \includegraphics[width=1\linewidth]{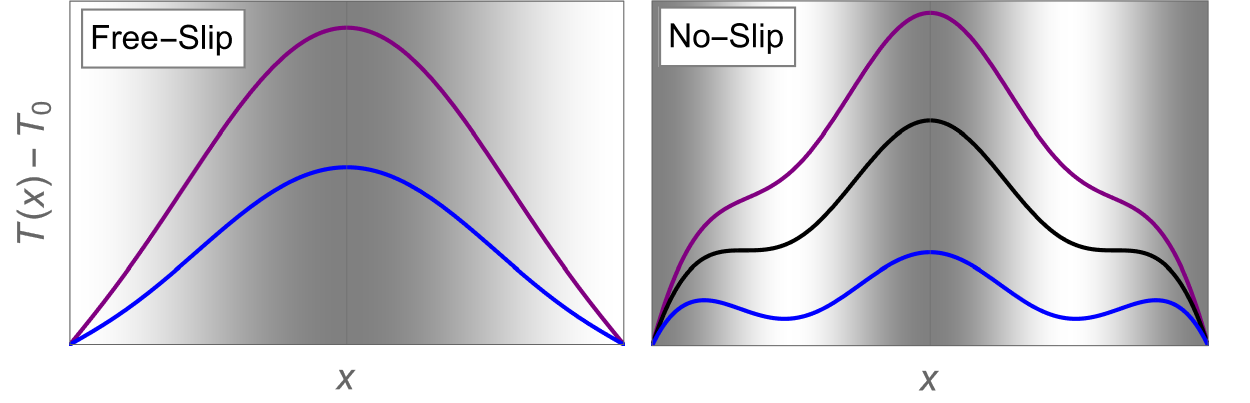}
    \caption{Temperature profiles near the spontaneous flow transition for varying $L_\epsilon$ (decreasing top to bottom). Dark bands depict regions of high shear. With no-slip boundary condition on the velocity, the temperature profile transitions from single peak to three peaks around $L_\epsilon \approx 0.875 L$.}
    \label{fig:Tprofile}
\end{figure}

The temperature profile depends qualitatively on the energy relaxation length $L_\epsilon$. Firstly, larger $L_\epsilon$ means that the energy gained from inhomogeneous shearing and twisting effects is transported over longer distances before relaxing to the environment, thereby heating the fluid more across the slab; see \cref{fig:Tprofile}. In the free-slip case, heating happens near the middle of the slab, so $T$ peaks in the middle and monotonically decreases to $T_0$ at the edges, with the overall amplitude controlled by $L_\epsilon$. However, in the no-slip case, heating happens both near the middle and edges of the slab. For $L_\epsilon > 0.875 L$, we find that the energy relaxes sufficiently slowly that the spatial heating structure is course-grained and, again, $T$ only peaks in the middle. However, for $L_\epsilon < 0.875 L$, the spatial heating structure dominates, resulting in two additional peaks near the edges.
By contrast, we show in Appendix~\ref{app:temperature-profile} that when we subject the temperature to Neumann boundary conditions, changing $L_{\epsilon}$ does not induce any qualitative change on the temperature profile. 

\vspace{0.5em}
\noindent
\textit{Discussion and perspectives---}%
In this work we used a novel hydrodynamic framework for active matter coupled to an environment to study the local temperature dynamics of a two-dimensional active nematic. This framework is based on the usual assumptions underlying hydrodynamics, namely local thermal equilibrium and microscopic reversibility \cite{grootmazur,J_licher_2018} but includes the coupling to an external environment \cite{Armas:2024iuy}. The interplay between energy inflow caused by microscopic self-driven agents and the outflow of energy due to exchanges with the environment allows for the existence of non-equilibrium steady states. 

We found that the mechanosensitive nature of the fuel consumption makes it so that the activity of active nematics does not linearly affect energy balance. However, when studying the well known spontaneous flow transition for a confined active nematic, we find that this does leave an imprint on the temperature profile. The nature of the temperature profile depends on the ratio between the diffusion along the confined geometry and the relaxation to the environment. It would be interesting to explore whether such temperature profiles could be observed in an experimental realization of a confined active nematic \cite{duclos2018spontaneous,duclos2}.

\vspace{1em}

\acknowledgments
We would like to thank Amin Doostmohammadi, Luca Giomi, Kristan Jensen, Pavel Kovtun, Benoit Mahault and Jonas Veenstra for helpful discussions. The authors are partly supported by the Dutch Institute for Emergent Phenomena (DIEP) cluster at the University of Amsterdam and JA via the DIEP programme Foundations and Applications of Emergence (FAEME).

\iffalse 
\begin{acknowledgments}
\noindent
\textit{Acknowledgments---}%

\end{acknowledgments}
\fi 


\clearpage
\appendix 

\onecolumngrid

\renewcommand{\thesection}{\Alph{section}}
\renewcommand{\thesubsection}{\thesection.\arabic{subsection}}
\renewcommand{\thesubsubsection}{\thesubsection.\arabic{subsubsection}}
% Fix references
\makeatletter
\renewcommand{\p@section}{}
\renewcommand{\p@subsection}{}
\renewcommand{\p@subsubsection}{}
\makeatother

\counterwithin{figure}{section}
\renewcommand\thefigure{\Alph{section}.\arabic{figure}}
\renewcommand{\theequation}{\Alph{section}.\arabic{equation}}

{\center{\large\bfseries Supplementary Material}\par}

\section{Dissipation rate and fluctuation-dissipation theorem}
\label{app:noisevariance}

In this section, we derive the the active nematic dissipation rate that accounts for entropy exchanges with the environment. We further obtain the distribution of stochastic noise fields for active nematics by using FDT near thermal equilibrium. Since we consider the active steady state to be close to thermal equilibrium, we can assume that the thermodynamic identities hold locally, i.e.
\begin{align}
    \df\varepsilon  &=  T \df s  + \mu\df\rho + \df\cF_\nem~, \qquad 
    \varepsilon = T s + \mu\rho - p + \cF_\nem~,
\end{align}
where $\mathcal{F}_\nem$ denotes the nematic free-energy contribution in \cref{eq:nematic-F}. Using this, together with the conservation equations, we can show that
\begin{align}
    T\dow_t s  &= \dow_t\epsilon
    - v_i\dow_t\pi^i
    - \lb\mu - \half v^2\rb\dow_t\rho
    + \cH^{ij}\dow_t Q_{ij}
    -  K \dow_i\!\lb \dow^iQ^{jk}\dow_t Q_{jk}\rb~.
\end{align}
Using equations of motion and conservation equations, we find
\begin{align}
    T\Theta_\tot \equiv T \Big(\dow_t s + \dow_i s^i + \kb r_\env\Big)
    &= 
    r_\fuel \Delta\mu + r_\env \kb \Delta T
    - \varepsilon^i_{\text{diss}}\frac1T\dow_i T 
    + t^{ij}_{\text{diss}} u_{ij} 
    + V^{ij}_{\text{diss}}\cH_{ij}~,
\end{align}
where we have identified the heat flux
\begin{align}
    Ts^i = \varepsilon^i + (p - \cF_\nem - \rho\mu) v^i 
    + K\dow^iQ^{jk}(\dow_t+v^l\dow_l) Q_{jk}~,
\end{align}
and isolated the dissipative part of the constitutive relations via
\begin{align}
    \varepsilon^i 
    &= \varepsilon v^i 
    - K\dow^iQ^{jk}(\dow_t+v^l\dow_l) Q_{jk}~, \nn\\
    t^{ij} 
    &= - \rho v^i v^j - p\delta^{ij} 
    + \cF_{\nem}\delta^{ij}
    - K\dow^iQ^{kl}\dow^j Q_{kl}
    + Q^i_{~k} \cH^{jk} - Q^j_{~k} \cH^{ik}
    + t^{ij}_{\text{diss}}~, \nn\\
    (\dow_t+v^k\dow_k) Q^{ij}
    &= - \Big( \Omega^{ i k}  Q_{k}^{\, \, j } - Q^i_{ \, \, k} \Omega^{k j} \Big) + V^{ij}_{\text{diss}}~.
\end{align}

The dissipation rate has the schematic form~\cite{Basu2008,J_licher_2018}
\begin{align}
    T \Theta_\tot  = \sum_{\alpha }   f_{\alpha} J^{\text{diss}}_{\alpha}~, 
\end{align}
where we have collectively represented the nonhydrostatic fluxes and the ``conjugate forces'', i.e.
\begin{align}
    J_\alpha^{\text{diss}}
    = \begin{pmatrix}
        r_\fuel \\ r_\env \\ \varepsilon^i_\diss \\
        t^{ij}_\diss \\ V^{ij}_\diss 
    \end{pmatrix}~, \qquad
    f_\alpha
    = \begin{pmatrix}
        \Delta\mu \\ \kb\Delta T \\ -\frac{1}{T}\dow_iT \\
        u_{ij} \\ \cH_{ij}
    \end{pmatrix}~.
\end{align}
In accordance with the second law of thermodynamics, the constitutive relations take the form
\begin{align}  \label{eq:constequations}
    J^\diss_{\alpha} = \sum_{\beta }O_{\alpha \beta } f_{\beta }
    + \xi_{\alpha}~~ . 
\end{align}
where $O_{\alpha \beta }$ is the Onsager matrix and $\xi_\alpha$ represent the noise fluxes. Comparing with the main text, the Onsager matrix takes the form
\begin{align}
    O_{\alpha \beta }
    = \begin{pmatrix}
        \gamma_\env & 0 & 0 & 0 & 0 \\
        0 & \gamma_\fuel & 0 & \alpha_{\fuel}    Q^{kl} & 0 \\
        0 & 0 & T\kappa & 0 & 0 \\
        0 &  - \alpha_{\fuel}'        Q^{ij} & 0 & 2\eta & - S\lambda \cL^{klij} \\
        0 & 0 & 0 & S\lambda' \cL^{ijkl} & \frac{1}{\sigma_Q} \cL^{ijkl}
    \end{pmatrix}~~,
\end{align}
where we have defined
\begin{align}
    \cL^{ijkl} = \half \Big( \delta^{ik}\delta^{jl} 
    + \delta^{il}\delta^{jk}
    - \delta^{ij}\delta^{kl}
    \Big)~~.
\end{align}
We have not introduced any off-diagonal mechanosensitive terms in $r_\env$ for simplicity.
Note that only the symmetric part of the matrix actually contributes to entropy production. Therefore, $\Theta_\env\geq 0$ requires that all eigenvalues of the symmetric Onsager matrix $\cO_{\alpha\beta}+\cO_{\beta\alpha}$ are non-negative.

Onsager reciprocity relations require that the Onsager matrix satisfy the symmetry property
\begin{align}
    O_{\beta\alpha} = \big({\mathbb T}O{\mathbb T}\big)_{\beta\alpha}~,
\end{align}
where ${\mathbb T}_{\alpha\beta}=\diag(-1,-1,-1,+1,-1)$ is the matrix of time-reversal eigenvalues of fluxes. This forces us to set the off-diagonal coefficients $\alpha'_{\fuel}=\alpha_{\fuel}$ and $\lambda'=\lambda$.

FDT dictates that the noise variance of the stochastic fluxes is given by \cite{PhysRev.91.1512,Fox1970,Forster2018}
\begin{align}
\Big\langle \xi_{\alpha}(x,t)\,\xi_{\beta }(x',t')\Big\rangle  
= \kb T_0 \big( O_{\alpha \beta}  + 
 O_{\beta \alpha } \big)\, \delta^{(2)}(x-x')\delta(t-t')~,
\end{align}
when fluctuating around a state with $\Delta\mu,\Delta T=0$. From here, we can read off the noise variances
\begin{align}\label{eq:noisevariances}
    \Big\langle \xi^i_{\epsilon} (x,t)\,\xi^j_{\epsilon}(x',t') \Big\rangle   
    &=  2 \kb 
   T_0^2 \kappa\, \delta^{ij} \delta^{(2)} (x-x')\delta(t-t')~, \nn\\ 
   \Big\langle  \xi^{ij}_{\pi}(x,t) \,\xi^{k l }_{\pi}(x',t') \Big\rangle 
   &=  2 \kb T_0  \eta\,
 \Big( \delta^{i   k } \delta^{lj} + \delta^{il} \delta^{kj} \Big)  \delta^{(2)} (x - x' )  \delta (t - t' )~, \nn\\
 \Big\langle \xi_{\fuel}(x,t)\,\xi_{\fuel}(x',t')\Big\rangle   
&=  2\kb T_0 \gamma_{\fuel}  \delta^{(2)}(x-x')\delta (t - t' )~, \nn\\ 
\Big\langle  \xi_{\env}(x,t) \,   \xi_{\env}(x',t') \Big\rangle  &  =  2  \kb  T_0      \gamma_{\env } \delta^{(2)} (x - x' )  \delta (t - t' )~, \nn\\
\Big\langle  \xi^{ij}_{Q}(x,t) \,\xi^{k l }_{Q}(x',t') \Big\rangle 
   &=  \frac{2 \kb T_0}{\sigma_Q}  \cL^{ijkl}  \delta^{(2)} (x - x' )  \delta (t - t' )~,
\end{align}
while all noise cross-correlations are zero.

\section{Linear instability of the homogeneous state in active nematics}
\label{sec:linear-instability}

In this appendix, we review the linearized instability of homogeneous steady states in active nematics. The homogeneous state given by $T=T_0$, $v^i=0$, $S=S_0$, and $\theta=\theta_0$. We can reach any state with $v^i\neq0$ by performing a Galilean boost transformation. We consider linearized plane-wave perturbations around this state $\propto\exp(-i\omega t + ik_i x^i)$. The linearized hydrodynamic equations yield
\begin{align}
    i\omega c_v\delta T
    - ik\delta\varepsilon_\|
    + t_{\perp\|}^{0} ik\delta v_\perp
    &= - r_\fuel \Delta \mu
    + r_\env  \kb  T_\env~, \nn\\
    i\omega \delta v_\perp +  \frac{ik}{\rho_0} \delta t_{\perp\|} 
    &= 0~, \nn\\
    i\omega\delta S
    + \lambda S_0 s_\theta ik \delta v_\perp
    &= -\frac{2}{\sigma_Q} \Big( c_\theta \cH_{\|\|} + s_\theta\cH_{\|\perp}\Big)
    ~,  \nn\\
    i\omega\delta\theta
    + \frac{1+\lambda c_\theta}{2} ik \delta v_\perp
    &= \frac{1}{S_0 \sigma_Q} \Big(s_\theta \cH_{\|\|} - c_\theta\cH_{\|\perp}\Big)~.
\end{align}
Here $\perp$ and $\|$ denote the components transverse and parallel to $k_i$ respectively. For a vector $X^i$, these are defined as $X_\| = k_i X^i/|k|$ and $X_\perp = k_i \epsilon^{ij}X_j/|k|$. Note that $\delta v_\|=0$ due to incompressibility. 
We have denoted $\cos(2\Delta\theta) = c_\theta$ and $\sin(2\Delta\theta) = s_\theta$ for brevity, where $\Delta\theta = \theta - \arctan(k_y/k_x)$ is the relative angle between the nematic orientation and the wavevector. In particular, $Q_{\|\|} = S/2\,c_\theta$ and $Q_{\|\perp} = S/2\,s_\theta$.
Various constitutive quantities appearing above are given by
\begin{align}
    \delta\varepsilon_\|
    &= - ik c_v D_\epsilon \delta T~, \nn\\
    \delta t_{\perp\|}
    &= 
    ik\eta\,\delta v_\perp 
    + S_0\Big(s_\theta\mathcal{H}_{\|\|} - c_\theta\mathcal{H}_{\|\perp}\Big)
    - \lambda S_0\mathcal{H}_{\|\perp}
     - \half \zeta   s_\theta\delta S
      - \zeta   S_0 c_\theta\delta\theta
     ~,  \nn\\
    \cH_{\|\|}
    &= K k^2 S_0 s_\theta\delta\theta
    - \lb \half K k^2 + b S_0^2 \rb c_\theta\delta S~, \nn\\
    \cH_{\|\perp}
    &= - K k^2 S_0 c_\theta\delta\theta
    - \lb \half Kk^2 + b S_0^2\rb s_\theta\delta S~,
\end{align}
together with $t_{\perp\|}^{0} = -  \half S_0 \zeta  s_\theta$, and the rate terms
\begin{align}
    r_\fuel \Delta\mu - r_\env \kb T_\env
    &= 
    - c_v\Gamma_\epsilon\delta T
     + \half S_0 \zeta  s_\theta ik \delta v_\perp~.
\end{align}

As argued in the main text, the temperature fluctuations decouple and give rise to simple diffusional dynamics
\begin{align}
    \lb i\omega - k^2 D_\epsilon - \Gamma_\epsilon \rb \delta T &= 0~.
\end{align}
The remaining three degrees of freedom give rise to
\begin{align}
    \begin{pmatrix}
        i\omega - D_\pi k^2 
        & - \frac{1}{\rho_0}\lb 1 + \lambda c_\theta\rb S_0^2 K k^4
        +  \frac{1}{\rho_0} c_\theta  \zeta  S_0 k^2
        & - \frac{1}{2\rho_0} s_\theta\lb \lambda S_0\lb Kk^2 + 2b S_0^2\rb 
        - \zeta  \rb k^2
        \\ 
        \half \lb 1+\lambda c_\theta\rb 
        & i\omega - D_Qk^2 & 0 \\
        S_0\lambda s_\theta & 0 & i\omega - D_Q k^2 - 2b S_0^2/\sigma_Q
    \end{pmatrix}
    \begin{pmatrix}
        ik \delta v_\perp \\ \delta\theta \\ \delta S
    \end{pmatrix}
    = 0~,
\end{align}
where $D_\pi = \eta/\rho_0$, $D_Q = K/\sigma_Q$. The linearized mode spectrum is obtained by setting the determinant of the matrix above to zero, i.e.
\begin{align}
    &\lb i\omega - D_Q k^2 - \frac{2b S_0^2}{\sigma_Q}\rb 
    \lb \lb i\omega - D_\pi k^2
    - \frac{\lambda^2 s_\theta^2 }{2\rho_0} S_0^2\sigma_Q k^2
    \rb
    \lb i\omega - D_Qk^2\rb 
    + \frac{1+\lambda c_\theta}{2\rho_0}S_0^2
    \lb \lb 1 + \lambda c_\theta\rb K k^2
    -  \frac{c_\theta  \zeta }{S_0} \rb k^2
    \rb \nn\\
    &\qquad
    + \frac{\lambda s_\theta^2}{2\rho_0} S_0^2\lb i\omega - D_Qk^2\rb
    \lb i\omega\sigma_Q\lambda
        -  \frac{ \zeta }{S_0} \rb k^2
    = 0.
\end{align}
In general, the roots of this equation are quite complicated. However, note that $\delta S$ fluctuations are gapped with relaxation rate $2bS_0^2/\sigma_Q$. We can focus on the low-energy regime, characterized by $|\omega| \ll 2bS_0^2/\sigma_Q$ and $k^2 \ll 2b S_0^2/K$, or, equivalently, send $b\to\infty$ in the above equation. This leaves us with quadratic dispersion relations for $\delta v_\perp$ and $\delta\theta$, taking the form
\begin{align}
    \lb i\omega - D_\pi k^2
    - \frac{\lambda^2 s_\theta^2 }{2\rho_0} S_0^2\sigma_Q k^2
    \rb
    \lb i\omega - D_Qk^2\rb 
    + \frac{1+\lambda c_\theta}{2\rho_0}S_0^2
    \lb \lb 1 + \lambda c_\theta\rb K k^2
    - \frac{c_\theta   \zeta }{S_0} \rb k^2 = 0.
\end{align}
This leads to two low-energy modes $\omega=-i\omega_\pm$, such that
\begin{align}
    \omega_++\omega_- 
    &= \lb D_Q + D_\pi
    + \frac{\lambda^2 s_\theta^2 }{2\rho_0} S_0^2\sigma_Q \rb  k^2~, \nn\\
    \omega_+\omega_-
    &= \lb D_Q D_\pi
    + \frac{1+\lambda^2+2\lambda c_\theta}{2\rho_0} S_0^2K
    \rb k^4
    - \frac{(1+\lambda c_\theta)c_\theta}{2\rho_0}S_0 \zeta  k^2~.
\end{align}
Due to the second law constraints, the sum of $\omega_\pm$ is decisively non-negative. As a consequence, we only need to require that the product of $\omega_\pm$ is non-negative to ensure that both the modes are stable and reside in the lower-half complex-$\omega$ plane. To wit, the stability criterion is
\begin{align}
    k^2
    \geq \frac{S_0 \zeta }{D_Q}
    \frac{(1+\lambda c_\theta)c_\theta}{
    2\eta + S_0^2\sigma_Q\big(1+\lambda^2+2\lambda c_\theta\big)}
    \equiv k_c(\Delta\theta)^2~.
\end{align}
Wave vectors smaller than this bound are unstable and precisely lead to the critical wavevector in \cref{eq:stability-bound}. This implies that the homogeneous state is unstable for $\cos(2\Delta\theta)<0$ for a contractile active nematic ($ \zeta  < 0$), peaking at $\Delta\theta=\pm\pi/2$ when the perturbations are transverse to the nematic orientation~\cite{PhysRevLett.89.058101}. Similarly, for an extensile active nematic ($ \zeta  > 0$), the instability arises for $\cos(2\Delta\theta)>0$, peaking at $\pm\Delta\theta=0,\pi$ when perturbations are along the nematic orientation. This is why the homogeneous state can be made stable by confining contractile active nematics perpendicular to the nematic order $L_\perp <\pi/k_c(\pi/2)$, while confining extensile active nematics parallel to the nematic order $L_\| <\pi/k_c(0)$~\cite{Giomi_2012, giomiroyal, Ramaswamy_2007, Voituriez_2005, 2024arXiv240316841L}.

\section{Perturbative computation of nematic flow transition}
\label{app:confinedgeometry}

In this appendix, we review the perturbative inhomogeneous steady state solution near the nematic flow transition. Given the symmetry of the problem, we assume that the solution only depends on the $x$-coordinate, as well as $v_x(x)=0$. Consequently, the momentum conservation equation in the $y$-direction reads $\dow_x t_{yx}(x)=0$, which can be solved to give $t_{yx}(x)=\sigma_0$ for some constant transverse stress $\sigma_0$. For free-slip boundary conditions, $t_{yx}(\pm L/2)=0$, we have $\sigma_0=0$. Whereas for no-slip boundary conditions, $v_y(\pm L/2)=0$, the constant $\sigma_0$ can be arbitrary.
This yields
\begin{align}
    S \sin(2\theta)\cH_{xx} 
    - S \cos(2\theta) \mathcal{H}_{xy}
    - S\lambda\mathcal{H}_{xy}
    -  \frac{S}{2} \zeta  \sin(2\theta)
    = \sigma_0
    - \eta\, \dow_x v_y~,
\end{align}
where
\begin{align}
    \cH_{xx}
    &= \half K\dow_x^2 \Big(S \cos(2\theta)\Big)
    - \half \lb a + bS^2 \rb S \cos(2\theta)~, \nn\\
    \cH_{xy}
    &= \half K\dow_x^2 \Big(S \sin(2\theta)\Big)
    - \half \lb a + bS^2 \rb S \sin(2\theta)~.
\end{align}
On the other hand, the nematic equation \eqref{eq:Qequation} yields
\begin{subequations}\label{eq:cH-profile}
\begin{align} 
    \cos(2\theta)\cH_{xx} + \sin(2\theta)\cH_{xy}
    &= - \frac{S\sigma_Q}{2} \lambda\sin(2\theta)\,\dow_x v_y~, \label{eq:Q-eqn1}\\
    \sin(2\theta)\cH_{xx}
    - \cos(2\theta)\cH_{xy}
    &= \frac{S\sigma_Q}{2} 
    \lb1+\lambda \cos(2\theta)\rb \dow_x v_y~. \label{eq:Q-eqn2}
\end{align}
\end{subequations}

We are interested in finding the steady state solutions of this setup, with the nematic orientation anchored at the boundaries $\theta(\pm L/2)=\theta_0$. We take $\theta_0=\pi/2$ for contractile active nematics ($ \zeta  < 0$) and $\theta_0=0$ for extensile active nematics ($ \zeta  > 0$)~\cite{PhysRevLett.92.118101}. While a fully non-linear solution can be constructed using numeric methods~\cite{Giomi_2012,Edwards_2009,PhysRevE.76.031921}, we can find a perturbative analytic solution around the homogeneous state. To this end, we introduce the perturbative expansion of various fields
\begin{align}
    S(x) &= S_0 + A_\theta S^{(1)}(x) + A_\theta^2 S^{(2)}(x) + \ldots~, \nn\\
    \theta(x) &= \theta_0 + A_\theta \theta^{(1)}(x) + A_\theta^2 \theta^{(2)}(x) + \ldots~, \nn\\
    v_y(x) &= 0 + A_\theta v_y^{(1)}(x) + A_\theta^2 v_y^{(2)}(x) + \ldots~,
\end{align}
controlled by a small parameter $A_\theta$, which we identify with the amplitude of $\theta(x)$ tilt in the middle of the slab, i.e. $A_\theta = \theta(0)-\theta_0$. Since $t_{yx}$ is zero in the homogeneous state, we must have that $\sigma_0\sim A_\theta$ with no-slip boundary conditions. Accordingly, we will expand $\sigma_0= A_\theta \sigma_0^{(1)} + A_\theta^2 \sigma_0^{(2)} + \ldots$.

Truncating \cref{eq:Q-eqn1} at $\cO(A_\theta)$, we find that $S^{(1)}$ decouples and satisfies
\begin{align}
    D_Q\dow_x^2 S^{(1)}
    - \frac{2b S_0^2}{\sigma_Q} S^{(1)}
    &= 0~.
\end{align}
We can solve this with Dirichlet boundary conditions, $S^{(1)}(\pm L/2)=0$, resulting in the trivial solution $S^{(1)}=0$. The remaining equations for $\theta^{(1)}$ and $v_y^{(1)}$ take the form
\begin{align}
    \frac{L_c^2}{\pi^2} \dow_x^2\theta^{(1)}
    + \theta^{(1)}
    =  - \frac{\sigma_0^{(1)}}{S_0 \zeta  \cos(2\theta_0)}~, \qquad 
    \dow_x v_y^{(1)}
    &= - \frac{2D_Q}{1+\lambda \cos(2\theta_0)}\dow_x^2\theta^{(1)}~,
\end{align}
where the critical length $L_c$ is given by
\begin{align}
    L_c^2
    &= \frac{\pi^2}{k_c(\theta_0)^2}
    = \frac{\pi^2 D_Q}{S_0 \zeta }
    \frac{
    2\eta + S_0^2\sigma_Q(1+\lambda\cos(2\theta_0))^2}{\cos(2\theta_0)\big(1+\lambda\cos(2\theta_0)\big)}~,
\end{align}
where we have used the fact that $\cos(2\theta_0)=\pm 1$ depending on the sign of $ \zeta $.
We use Dirichlet boundary conditions for the nematic orientation, $\theta^{(1)}(\pm L/2)=0$. With free-slip boundary conditions for velocity, $t_{yx}(\pm L/2)=0$, which results in $\sigma_0^{(1)}=0$, a non-trivial steady state solution to the linearized equations can only be constructed when $L$ is a multiple of $L_c$. A steady state solution may exist at other arbitrary values of $L$, but requires us to invoke the fully non-linear hydrodynamic model~\cite{Giomi_2012,Edwards_2009,PhysRevE.76.031921}. The simplest such solution exists for $L\gtrsim L_c$~\cite{Voituriez_2005}, given by
\begin{align}\label{eq:sol-FS}
    \theta^{(1)}(x) \Big|_{\text{free-slip}}
    \approx  A_{\theta } \cos(\pi x/L)~, \qquad 
    v_y^{(1)} \Big|_{\text{free-slip}}
    \approx \frac{2\pi A_\theta D_Q}{\big(1+\lambda \cos(2\theta_0)\big)L}\sin(\pi x/L)
    + v^{(1)}_0~,
\end{align}
where $v^{(1)}_0$ is an arbitrary constant, which is set to zero by requiring that the average velocity along the slab is zero. Invoking non-linear corrections, one finds that the amplitude scales as $A_\theta\sim\sqrt{L-L_c}$.

On the other hand, with no-slip boundary conditions for velocity, $v_{y}(\pm L/2)=0$, a non-trivial linearized steady state solution only exists when $L$ is a multiple of $2L_c$. For $L\gtrsim 2L_c$, the solution is given by~\cite{Voituriez_2005}
\begin{align}\label{eq:sol-NS}
    \theta^{(1)} \Big|_{\text{no-slip}}
    &\approx A_\theta\cos^2(\pi x/L)
    - v^{(1)}_0 \frac{\big(1+\lambda \cos(2\theta_0)\big) L}{4\pi D_Q} 
    \sin(2\pi x/L)~, \nn\\ 
    v_y^{(1)} \Big|_{\text{no-slip}}
    &\approx \frac{2\pi A_\theta D_Q}{\big(1+\lambda \cos(2\theta_0)\big) L}
    \sin(2\pi x/L) 
    + 2v^{(1)}_0 \cos^2(\pi x/L)
    ~,
\end{align}
where we have identified $\sigma_0^{(1)}=  - \half A_\theta S_0  \zeta  \cos(2\theta_0)$. In this instance, the amplitude scales as $A_\theta\sim\sqrt{L-2L_c}$. The constant $v^{(1)}_0$ can again be set to zero by requiring that the average velocity along the slab is zero.

Using these solutions in \cref{eq:sol-FS,eq:sol-NS}, one may verify that the approximate shear profile near the spontaneous flow transition is given by \cref{eq:shear-profile} with 
\begin{align}
    A_u = \frac{\pi^2 A_\theta D_Q}{(1+\lambda\cos(2\theta_0)) L^2}~.
\end{align}

\section{Temperature profile of nematic flow transition}
\label{app:temperature-profile}

To compute the temperature profile of this solution, we invoke the energy balance equation \eqref{eq:energybalance123}, which results in
\begin{align}
    \kappa \partial_x^2 T
    + \frac{\dow\kappa}{\dow T}(\dow_x T)^2
    - c_v \Gamma_\epsilon (T-T_0)
    &=
    - \eta\, (\dow_x v_y)^2
    - \frac{2}{\sigma_Q } \lb\cH_{xx}^2 + \cH_{xy}^2\rb \nn\\
    &= 
    - \lb \eta
    + \frac{S^2\sigma_Q}{2} 
    \lb1+\lambda^2 + 2\lambda \cos(2\theta)\rb \rb (\dow_x v_y)^2~.
\end{align}
In the second line, we have used \cref{eq:cH-profile} to eliminate $\cH_{ij}$. To solve this equation perturbatively around the homogeneous state, we introduce the expansion
\begin{align}\label{eq:temperature-expansion}
    T(x) &= T_0 + A_\theta T^{(1)}(x) + A_\theta^2 T^{(2)}(x) + \ldots~.
\end{align}
At $\cO(A_\theta)$, the temperature fluctuations decouple and satisfy $D_\epsilon\partial_x^2T^{(1)} - \Gamma_\epsilon T^{(1)} = 0$. Solving these with Dirichlet boundary conditions $T(\pm L/2)=0$ or Neumann boundary conditions $\dow_x T(\pm L/2)=0$, we find the trivial profile $T^{(1)}(x)=0$. Hence, first non-trivial corrections to the temperature profile arise at $\cO(A_\theta^2)$, with the differential equation
\begin{align}\label{eq:DE-T2}
    D_\epsilon \partial_x^2 T^{(2)}
    - \Gamma_\epsilon T^{(2)}
    &= 
    - \frac{\tilde\eta}{c_v} \lb\dow_x v^{(1)}_y\rb^2~, \qquad 
    \tilde\eta = \eta
    + \frac{S_0^2\sigma_Q}{2} 
    \lb1+\lambda\cos(2\theta_0)\rb^2~.
\end{align}
Substituting the solution for $v^{(1)}_y$, this can be expressed as an inhomogeneous differential equation
\begin{align}\label{eq:inhomogeneousdifferentialequation123}
    \partial^2_x T^{(2)}(x)
    - k_\epsilon^2 T^{(2)}(x) 
    = - \Phi (x )~, 
\end{align}
where $k_\epsilon^2 = \Gamma_\epsilon/D_\epsilon$ and the source is given as
\begin{align}
    \Phi(x)\Big|_{\text{free-slip}}
    &= 
    \frac{4A_u^2\tilde\eta}{\kappa} \cos^2(\pi x/L)~, \qquad
    \Phi(x)\Big|_{\text{no-slip}}
    = \frac{16A_u^2\tilde\eta}{\kappa}\cos^2(2\pi x/L)~.
\end{align}

\subsection{Dirichlet boundary conditions}

Let us first solve for the temperature profile with Dirichlet boundary conditions $T^{(2)}(\pm L/2)=0$.
The differential equation can be solved with
\begin{align}\label{eq:T2result}
      T^{(2)}(x)  = \int^{L / 2 }_{- L/2 } \df x' G(x,x')\Phi(x')~,  
\end{align}
where $G(x,x')$ is the Green's function that satisfies
\begin{align}\label{eq:GreensFunction-DE}
     \Big( k^2_\epsilon - \partial^2_x \Big)
     G  (x , x' )  = \delta (x , x' ) ~~ . 
\end{align}
Imposition of Dirichlet boundary conditions at \( x,x' = \pm L/2 \) leads to
\begin{align}
G^D (x, x') =
\begin{cases}
\dsp\frac{\sinh\!\Big( k_{\epsilon}(L/2+x) \Big) \sinh\!\Big( k_{\epsilon} (L/2 - x') \Big)}{k_{\epsilon}\sinh(k_{\epsilon} L)}, & x < x'~, \\[10pt]
\dsp\frac{\sinh\!\Big( k_{\epsilon}(L/2+x') \Big) \sinh\!\Big( k_{\epsilon} (L/2 - x) \Big)}{k_{\epsilon}\sinh(k_{\epsilon} L)}, & x > x'~.
\end{cases}
\end{align}
We then plug this into \eqref{eq:T2result} to find the temperature profile
\begin{align}\label{eq:dirichletbdry}
    T^{(2)}(x)\Big|_{\text{free-slip}}
    &\approx \frac{L^2 A_u^2\tilde\eta/\kappa}{\pi ^2 (1+L^2/L_\epsilon^2)}
    \left(
    \frac{L_\epsilon^2}{2L^2}\lb 1
    - \frac{\cosh(2\pi x/L_\epsilon)}{\cosh(\pi L/L_\epsilon)} \rb
    + \cos^2(\pi x/L)
    \right)~, \nn\\
    T^{(2)}(x)\Big|_{\text{no-slip}}
    &\approx \frac{4L^2 A_u^2\tilde\eta/\kappa}{\pi^2(4+L^2/L_\epsilon^2)}
    \left(
    \lb 1 + \frac{2L_\epsilon^2}{L^2}\rb \lb
    1 - \frac{\cosh(2\pi x/L_\epsilon)}{\cosh(\pi L/L_\epsilon)} \rb
    - \sin^2(2\pi x/L) \right)~,
\end{align}
where $L_\epsilon = 2\pi/k_\epsilon$. We have recovered the results presented in the main text and arrived at the temperature profile in \cref{eq:T-profile}.

\subsection{Neumann boundary conditions}

\begin{figure}[t]
    \centering 
    \includegraphics[width=0.7\linewidth]{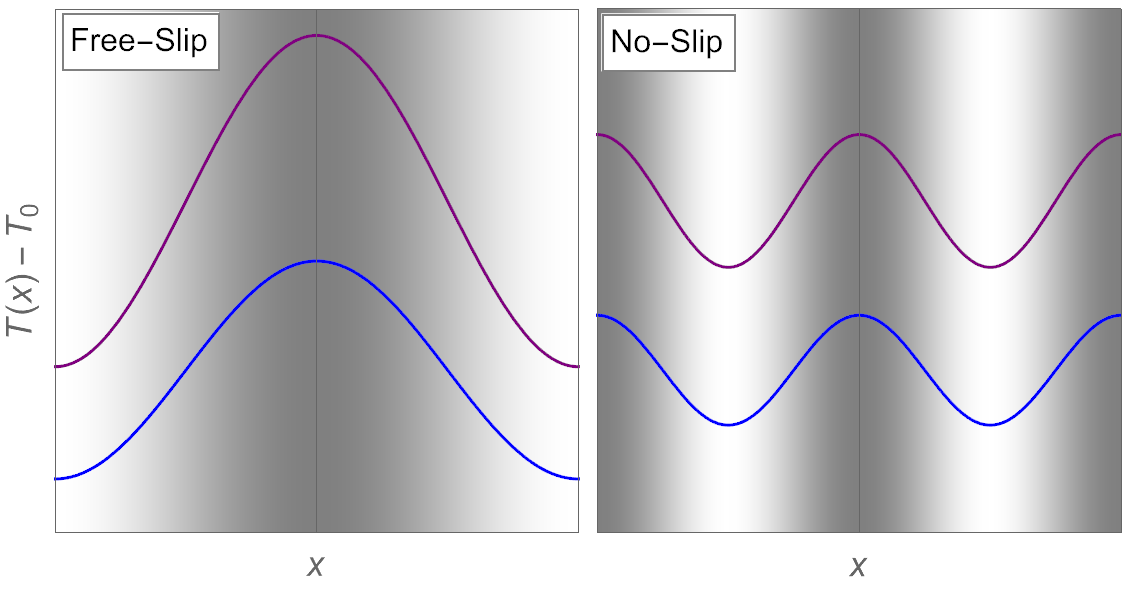}
    \caption{Temperature profiles near the spontaneous flow transition for varying $L_\epsilon$ (decreasing top to bottom). Dark bands depict regions of high shear.}
    \label{fig:TprofileN}
\end{figure}

Let us now look at the temperature profile when subjected to Neumann boundary conditions, $\dow_x T^{(2)}(\pm L/2)=0$. 
% Implementing the temperature expansion in \cref{eq:temperature-expansion}, we again find that $T^{(1)}(x)=0$, while the differential equation for $T^{(2)}(x)$ in \cref{eq:DE-T2} can be solved in terms of the the Green's function in \cref{eq:T2result}.
% \begin{align}  \label{eq:T2result-neumann}
%       T^{(2)}(x)  = \int^{L / 2 }_{- L/2 } \df x' G(x,x')\Phi(x')~,  
% \end{align}
% where $G(x,x')$ is the Green's function that satisfies
% \begin{align}
%      \Big( k^2_\epsilon - \partial^2_x \Big)
%      G  (x , x' )  = \delta (x , x' ) ~~ . 
% \end{align}
Imposing Neumann boundary conditions to the Green's function in \cref{eq:GreensFunction-DE} at \( x,x' = \pm L/2 \) leads to
\begin{align}
G^N(x, x') =
\begin{cases}
\dsp\frac{\cosh\!\Big( k_{\epsilon}(L/2+x) \Big) \cosh\!\Big( k_{\epsilon} (L/2 - x') \Big)}{k_{\epsilon}\sinh(k_{\epsilon} L)}, & x < x'~, \\[10pt]
\dsp\frac{\cosh\!\Big( k_{\epsilon}(L/2+x') \Big) \cosh\!\Big( k_{\epsilon} (L/2 - x) \Big)}{k_{\epsilon}\sinh(k_{\epsilon} L)}, & x > x'~.
\end{cases}
\end{align}
We then plug this into \eqref{eq:T2result} to find the temperature profile
\begin{align}\label{eq:neumannbdry}
    T^{(2)}(x)\Big|_{\text{free-slip}}
    &\approx \frac{L^2 A_u^2\tilde\eta/\kappa}{\pi ^2 (1+L^2/L_\epsilon^2)}
    \left(
    \frac{L_\epsilon^2}{2L^2}
    + \cos^2(\pi x/L)
    \right)~, \nn\\
    T^{(2)}(x)\Big|_{\text{no-slip}}
    &\approx \frac{4L^2 A_u^2\tilde\eta/\kappa}{\pi^2(4 +L^2/L_\epsilon^2)}
    \left(    \frac{2 L_\epsilon^2}{L^2}  +  \cos^2(2\pi x/L) \right)~,
\end{align}
where $L_\epsilon = 2\pi/k_\epsilon$. The resulting temperature profile is shown in Fig.~\ref{fig:TprofileN}. Note that unlike for the case with Dirichlet boundary conditions, the shape of the profile is not affected by $L_{\epsilon}$ for the case of Neumann boundary conditions. The reason for this is that Neumann boundary conditions force the energy current to be zero at the boundary, thus allowing temperature to be lost exclusively through thermal relaxation and not through diffusion. Therefore, the profile does not display the same competition between diffusive and relaxational energy loss that was found for the profile with Dirichlet boundary conditions studied in the main text.

\section{Parameter estimate for thermal modulations in a confined geometry} 
\label{app:estimate}

In this appendix, we attempt to estimate the strength of thermal oscillations for the spontaneous flow transition that arise for a cell monolayer in a confined geometry as observed in~\cite{duclos2018spontaneous}. Let us consider the profile for Neumann boundary conditions for simplicity. The magnitude of the thermal modulation for Neumann boundary conditions as given by \cref{eq:dirichletbdry,eq:neumannbdry} is of the form
\begin{align}
\Delta   T^{(2)} \sim   \frac{L^2 A_u^2\left( 
\eta
    + \frac{S_0^2\sigma_Q}{2} 
    \left( 1+\lambda \right)^2 \right)  /\kappa}{\pi ^2 (1+L^2/L_\epsilon^2)}~~.
\end{align}
Note that this estimate is $\propto 1 / \kappa$ because $\kappa$, unlike $\Gamma_{\epsilon }$, punishes the inhomogeneity of a thermal profile. Using \cite{duclos2018spontaneous}, we can estimate the values of some of the parameters for this confined geometry as reported below in Table 1.
\begin{table}[h!]
\centering
\begin{tabular}{ll}
\hline
\textbf{Parameter} & \textbf{ Value} \\
\hline
$\eta$ & $\sim 10^4~\text{Pa} \cdot \text{hr} \cdot \mu\text{m}$ \\
$L$ & $300~\mu\text{m}$ \\
$\kappa$ & $?$ \\
$\lambda$ & $\sim 1$ \\
$S_0$ & $\sim 1$ \\
$\sigma_Q$ & $\sim 10^4~\text{Pa} \cdot \text{hr} \cdot \mu\text{m}$ \\
$A_u$ & $ ?$ \\
$L_{\epsilon}$ & $?$ \\
\hline
\end{tabular}
\caption{Overview of model parameters and their estimated values based on \cite{duclos2018spontaneous}.}
\end{table}\\
The entries with a question mark are harder to estimate but are nevertheless important in order to provide an estimate of the modulation amplitude. Of special importance are those related to thermal flow and $A_u$, which is related to the bifurcation, and should be zero precisely at the critical point. However, we can infer $A_u$ from the maximum velocity of the confining free-slip geometry $v_{\text{max}} = 5\mu\text{m/hr}$. In particular we find the relation
\begin{align}
     v_{\text{max}} = \frac{2 L }{\pi } A_{u }  ~~ ,
\end{align}
so that we have access to $A_u$ and can be estimated to be
\begin{align}
    A_u \sim  0.03 \text{hr}^{-1} ~~.
\end{align}
For the thermal conductivity, we use the value characteristic of water since it is typically a good estimate for cellular matter, thus
\begin{align}
    \kappa_\text{water} \sim  0.6\,\text{W/(m$\cdot$K)}  = 2.16 \times 10^{15}
    \text{Pa} \cdot \mu\text{m}^2/(\text{hr}\cdot\text{K}) ~~.
\end{align}
The active nematic in \cite{duclos2018spontaneous} is a monolayer of cells, which has thickness of $h \sim   5 \mu m $. Thus, we have
\begin{align}
    \kappa =  \kappa_\text{water} h  \sim     10^{16} \text{Pa} \cdot \mu\text{m}^3/(\text{hr}\cdot\text{K})~~, 
\end{align}
in turn implying that 
\begin{align}
  \Delta T^{(2)} \sim    \frac{10^{-11}}{  1+L^2/L_\epsilon^2} {\text K}~~.
\end{align}
This heuristic estimate of temperature modulation currently appears too small to be measured directly. Nevertheless, in light of our results, it would be intriguing to explore whether this estimate could be increased to experimentally resolvable temperature scales by investigating the spontaneous flow transition across different physical realisations of active nematicity.

% It would be interesting to know whether such magnitude of temperature modulation, heuristically estimated here to be in the $\sim\text{mK}$ range, could be experimentally measured.


\begin{thebibliography}{61}%
\makeatletter
\providecommand \@ifxundefined [1]{%
 \@ifx{#1\undefined}
}%
\providecommand \@ifnum [1]{%
 \ifnum #1\expandafter \@firstoftwo
 \else \expandafter \@secondoftwo
 \fi
}%
\providecommand \@ifx [1]{%
 \ifx #1\expandafter \@firstoftwo
 \else \expandafter \@secondoftwo
 \fi
}%
\providecommand \natexlab [1]{#1}%
\providecommand \enquote  [1]{``#1''}%
\providecommand \bibnamefont  [1]{#1}%
\providecommand \bibfnamefont [1]{#1}%
\providecommand \citenamefont [1]{#1}%
\providecommand \href@noop [0]{\@secondoftwo}%
\providecommand \href [0]{\begingroup \@sanitize@url \@href}%
\providecommand \@href[1]{\@@startlink{#1}\@@href}%
\providecommand \@@href[1]{\endgroup#1\@@endlink}%
\providecommand \@sanitize@url [0]{\catcode `\\12\catcode `\$12\catcode `\&12\catcode `\#12\catcode `\^12\catcode `\_12\catcode `\%12\relax}%
\providecommand \@@startlink[1]{}%
\providecommand \@@endlink[0]{}%
\providecommand \url  [0]{\begingroup\@sanitize@url \@url }%
\providecommand \@url [1]{\endgroup\@href {#1}{\urlprefix }}%
\providecommand \urlprefix  [0]{URL }%
\providecommand \Eprint [0]{\href }%
\providecommand \doibase [0]{https://doi.org/}%
\providecommand \selectlanguage [0]{\@gobble}%
\providecommand \bibinfo  [0]{\@secondoftwo}%
\providecommand \bibfield  [0]{\@secondoftwo}%
\providecommand \translation [1]{[#1]}%
\providecommand \BibitemOpen [0]{}%
\providecommand \bibitemStop [0]{}%
\providecommand \bibitemNoStop [0]{.\EOS\space}%
\providecommand \EOS [0]{\spacefactor3000\relax}%
\providecommand \BibitemShut  [1]{\csname bibitem#1\endcsname}%
\let\auto@bib@innerbib\@empty
%</preamble>
\bibitem [{\citenamefont {Chaikin}\ and\ \citenamefont {Lubensky}(1995)}]{Chaikin_Lubensky_1995}%
  \BibitemOpen
  \bibfield  {author} {\bibinfo {author} {\bibfnamefont {P.~M.}\ \bibnamefont {Chaikin}}\ and\ \bibinfo {author} {\bibfnamefont {T.~C.}\ \bibnamefont {Lubensky}},\ }\href@noop {} {\emph {\bibinfo {title} {Principles of Condensed Matter Physics}}}\ (\bibinfo  {publisher} {Cambridge University Press},\ \bibinfo {year} {1995})\BibitemShut {NoStop}%
\bibitem [{\citenamefont {de~Gennes}\ and\ \citenamefont {Prost}(1993)}]{de1993physics}%
  \BibitemOpen
  \bibfield  {author} {\bibinfo {author} {\bibfnamefont {P.}~\bibnamefont {de~Gennes}}\ and\ \bibinfo {author} {\bibfnamefont {J.}~\bibnamefont {Prost}},\ }\href {https://books.google.nl/books?id=0Nw-dzWz5agC} {\emph {\bibinfo {title} {The Physics of Liquid Crystals}}},\ International Series of Monographs on Physics\ (\bibinfo  {publisher} {Clarendon Press},\ \bibinfo {year} {1993})\BibitemShut {NoStop}%
\bibitem [{\citenamefont {Doostmohammadi}\ \emph {et~al.}(2018)\citenamefont {Doostmohammadi}, \citenamefont {Ign{\'e}s-Mullol}, \citenamefont {Yeomans},\ and\ \citenamefont {Sagu{\'e}s}}]{Doostmohammadi2018}%
  \BibitemOpen
  \bibfield  {author} {\bibinfo {author} {\bibfnamefont {A.}~\bibnamefont {Doostmohammadi}}, \bibinfo {author} {\bibfnamefont {J.}~\bibnamefont {Ign{\'e}s-Mullol}}, \bibinfo {author} {\bibfnamefont {J.~M.}\ \bibnamefont {Yeomans}},\ and\ \bibinfo {author} {\bibfnamefont {F.}~\bibnamefont {Sagu{\'e}s}},\ }\bibfield  {title} {\bibinfo {title} {Active nematics},\ }\href {https://doi.org/10.1038/s41467-018-05666-8} {\bibfield  {journal} {\bibinfo  {journal} {Nature Communications}\ }\textbf {\bibinfo {volume} {9}},\ \bibinfo {pages} {3246} (\bibinfo {year} {2018})}\BibitemShut {NoStop}%
\bibitem [{\citenamefont {Saw}\ \emph {et~al.}(2017{\natexlab{a}})\citenamefont {Saw}, \citenamefont {Doostmohammadi}, \citenamefont {Nier}, \citenamefont {Kocgozlu}, \citenamefont {Thampi}, \citenamefont {Toyama}, \citenamefont {Marcq}, \citenamefont {Lim}, \citenamefont {Yeomans},\ and\ \citenamefont {Ladoux}}]{Saw2017}%
  \BibitemOpen
  \bibfield  {author} {\bibinfo {author} {\bibfnamefont {T.~B.}\ \bibnamefont {Saw}}, \bibinfo {author} {\bibfnamefont {A.}~\bibnamefont {Doostmohammadi}}, \bibinfo {author} {\bibfnamefont {V.}~\bibnamefont {Nier}}, \bibinfo {author} {\bibfnamefont {L.}~\bibnamefont {Kocgozlu}}, \bibinfo {author} {\bibfnamefont {S.}~\bibnamefont {Thampi}}, \bibinfo {author} {\bibfnamefont {Y.}~\bibnamefont {Toyama}}, \bibinfo {author} {\bibfnamefont {P.}~\bibnamefont {Marcq}}, \bibinfo {author} {\bibfnamefont {C.~T.}\ \bibnamefont {Lim}}, \bibinfo {author} {\bibfnamefont {J.~M.}\ \bibnamefont {Yeomans}},\ and\ \bibinfo {author} {\bibfnamefont {B.}~\bibnamefont {Ladoux}},\ }\bibfield  {title} {\bibinfo {title} {Topological defects in epithelia govern cell death and extrusion},\ }\href {https://doi.org/10.1038/nature21718} {\bibfield  {journal} {\bibinfo  {journal} {Nature}\ }\textbf {\bibinfo {volume} {544}},\ \bibinfo {pages} {212} (\bibinfo {year} {2017}{\natexlab{a}})}\BibitemShut {NoStop}%
\bibitem [{\citenamefont {Assante}\ \emph {et~al.}(2023)\citenamefont {Assante}, \citenamefont {Corbett}, \citenamefont {Marenduzzo},\ and\ \citenamefont {Morozov}}]{Assante2023}%
  \BibitemOpen
  \bibfield  {author} {\bibinfo {author} {\bibfnamefont {R.}~\bibnamefont {Assante}}, \bibinfo {author} {\bibfnamefont {D.}~\bibnamefont {Corbett}}, \bibinfo {author} {\bibfnamefont {D.}~\bibnamefont {Marenduzzo}},\ and\ \bibinfo {author} {\bibfnamefont {A.}~\bibnamefont {Morozov}},\ }\bibfield  {title} {\bibinfo {title} {Active turbulence and spontaneous phase separation in inhomogeneous extensile active gels},\ }\href {https://doi.org/10.1039/d2sm01188c} {\bibfield  {journal} {\bibinfo  {journal} {Soft Matter}\ }\textbf {\bibinfo {volume} {19}},\ \bibinfo {pages} {189–198} (\bibinfo {year} {2023})}\BibitemShut {NoStop}%
\bibitem [{\citenamefont {Marchetti}\ \emph {et~al.}(2013)\citenamefont {Marchetti}, \citenamefont {Joanny}, \citenamefont {Ramaswamy}, \citenamefont {Liverpool}, \citenamefont {Prost}, \citenamefont {Rao},\ and\ \citenamefont {Simha}}]{RevModPhys.85.1143}%
  \BibitemOpen
  \bibfield  {author} {\bibinfo {author} {\bibfnamefont {M.~C.}\ \bibnamefont {Marchetti}}, \bibinfo {author} {\bibfnamefont {J.~F.}\ \bibnamefont {Joanny}}, \bibinfo {author} {\bibfnamefont {S.}~\bibnamefont {Ramaswamy}}, \bibinfo {author} {\bibfnamefont {T.~B.}\ \bibnamefont {Liverpool}}, \bibinfo {author} {\bibfnamefont {J.}~\bibnamefont {Prost}}, \bibinfo {author} {\bibfnamefont {M.}~\bibnamefont {Rao}},\ and\ \bibinfo {author} {\bibfnamefont {R.~A.}\ \bibnamefont {Simha}},\ }\bibfield  {title} {\bibinfo {title} {Hydrodynamics of soft active matter},\ }\href {https://doi.org/10.1103/RevModPhys.85.1143} {\bibfield  {journal} {\bibinfo  {journal} {Rev. Mod. Phys.}\ }\textbf {\bibinfo {volume} {85}},\ \bibinfo {pages} {1143} (\bibinfo {year} {2013})}\BibitemShut {NoStop}%
\bibitem [{\citenamefont {van Saarloos}\ \emph {et~al.}(2024)\citenamefont {van Saarloos}, \citenamefont {Vitelli},\ and\ \citenamefont {Zeravcic}}]{Van_Saarloos2024}%
  \BibitemOpen
  \bibfield  {author} {\bibinfo {author} {\bibfnamefont {W.}~\bibnamefont {van Saarloos}}, \bibinfo {author} {\bibfnamefont {V.}~\bibnamefont {Vitelli}},\ and\ \bibinfo {author} {\bibfnamefont {Z.}~\bibnamefont {Zeravcic}},\ }\href@noop {} {\emph {\bibinfo {title} {Soft matter}}}\ (\bibinfo  {publisher} {Princeton University Press},\ \bibinfo {address} {Princeton, NJ},\ \bibinfo {year} {2024})\BibitemShut {NoStop}%
\bibitem [{\citenamefont {Sanchez}\ \emph {et~al.}(2012)\citenamefont {Sanchez}, \citenamefont {Chen}, \citenamefont {DeCamp}, \citenamefont {Heymann},\ and\ \citenamefont {Dogic}}]{Sanchez2012}%
  \BibitemOpen
  \bibfield  {author} {\bibinfo {author} {\bibfnamefont {T.}~\bibnamefont {Sanchez}}, \bibinfo {author} {\bibfnamefont {D.~T.~N.}\ \bibnamefont {Chen}}, \bibinfo {author} {\bibfnamefont {S.~J.}\ \bibnamefont {DeCamp}}, \bibinfo {author} {\bibfnamefont {M.}~\bibnamefont {Heymann}},\ and\ \bibinfo {author} {\bibfnamefont {Z.}~\bibnamefont {Dogic}},\ }\bibfield  {title} {\bibinfo {title} {Spontaneous motion in hierarchically assembled active matter},\ }\href {https://doi.org/10.1038/nature11591} {\bibfield  {journal} {\bibinfo  {journal} {Nature}\ }\textbf {\bibinfo {volume} {491}},\ \bibinfo {pages} {431} (\bibinfo {year} {2012})}\BibitemShut {NoStop}%
\bibitem [{\citenamefont {Giomi}\ \emph {et~al.}(2013)\citenamefont {Giomi}, \citenamefont {Bowick}, \citenamefont {Ma},\ and\ \citenamefont {Marchetti}}]{PhysRevLett.110.228101}%
  \BibitemOpen
  \bibfield  {author} {\bibinfo {author} {\bibfnamefont {L.}~\bibnamefont {Giomi}}, \bibinfo {author} {\bibfnamefont {M.~J.}\ \bibnamefont {Bowick}}, \bibinfo {author} {\bibfnamefont {X.}~\bibnamefont {Ma}},\ and\ \bibinfo {author} {\bibfnamefont {M.~C.}\ \bibnamefont {Marchetti}},\ }\bibfield  {title} {\bibinfo {title} {Defect annihilation and proliferation in active nematics},\ }\href {https://doi.org/10.1103/PhysRevLett.110.228101} {\bibfield  {journal} {\bibinfo  {journal} {Phys. Rev. Lett.}\ }\textbf {\bibinfo {volume} {110}},\ \bibinfo {pages} {228101} (\bibinfo {year} {2013})}\BibitemShut {NoStop}%
\bibitem [{\citenamefont {Alert}\ \emph {et~al.}(2020)\citenamefont {Alert}, \citenamefont {Joanny},\ and\ \citenamefont {Casademunt}}]{Alert_2020}%
  \BibitemOpen
  \bibfield  {author} {\bibinfo {author} {\bibfnamefont {R.}~\bibnamefont {Alert}}, \bibinfo {author} {\bibfnamefont {J.-F.}\ \bibnamefont {Joanny}},\ and\ \bibinfo {author} {\bibfnamefont {J.}~\bibnamefont {Casademunt}},\ }\bibfield  {title} {\bibinfo {title} {Universal scaling of active nematic turbulence},\ }\href {https://doi.org/10.1038/s41567-020-0854-4} {\bibfield  {journal} {\bibinfo  {journal} {Nature Physics}\ }\textbf {\bibinfo {volume} {16}},\ \bibinfo {pages} {682–688} (\bibinfo {year} {2020})}\BibitemShut {NoStop}%
\bibitem [{\citenamefont {Alert}\ \emph {et~al.}(2022)\citenamefont {Alert}, \citenamefont {Casademunt},\ and\ \citenamefont {Joanny}}]{turbulencealert}%
  \BibitemOpen
  \bibfield  {author} {\bibinfo {author} {\bibfnamefont {R.}~\bibnamefont {Alert}}, \bibinfo {author} {\bibfnamefont {J.}~\bibnamefont {Casademunt}},\ and\ \bibinfo {author} {\bibfnamefont {J.-F.}\ \bibnamefont {Joanny}},\ }\bibfield  {title} {\bibinfo {title} {Active turbulence},\ }\href {https://doi.org/https://doi.org/10.1146/annurev-conmatphys-082321-035957} {\bibfield  {journal} {\bibinfo  {journal} {Annual Review of Condensed Matter Physics}\ }\textbf {\bibinfo {volume} {13}},\ \bibinfo {pages} {143} (\bibinfo {year} {2022})}\BibitemShut {NoStop}%
\bibitem [{\citenamefont {Peruani}\ \emph {et~al.}(2012)\citenamefont {Peruani}, \citenamefont {Starru\ss{}}, \citenamefont {Jakovljevic}, \citenamefont {S\o{}gaard-Andersen}, \citenamefont {Deutsch},\ and\ \citenamefont {B\"ar}}]{PhysRevLett.108.098102}%
  \BibitemOpen
  \bibfield  {author} {\bibinfo {author} {\bibfnamefont {F.}~\bibnamefont {Peruani}}, \bibinfo {author} {\bibfnamefont {J.}~\bibnamefont {Starru\ss{}}}, \bibinfo {author} {\bibfnamefont {V.}~\bibnamefont {Jakovljevic}}, \bibinfo {author} {\bibfnamefont {L.}~\bibnamefont {S\o{}gaard-Andersen}}, \bibinfo {author} {\bibfnamefont {A.}~\bibnamefont {Deutsch}},\ and\ \bibinfo {author} {\bibfnamefont {M.}~\bibnamefont {B\"ar}},\ }\bibfield  {title} {\bibinfo {title} {Collective motion and nonequilibrium cluster formation in colonies of gliding bacteria},\ }\href {https://doi.org/10.1103/PhysRevLett.108.098102} {\bibfield  {journal} {\bibinfo  {journal} {Phys. Rev. Lett.}\ }\textbf {\bibinfo {volume} {108}},\ \bibinfo {pages} {098102} (\bibinfo {year} {2012})}\BibitemShut {NoStop}%
\bibitem [{\citenamefont {Gro\ss{}mann}\ \emph {et~al.}(2016)\citenamefont {Gro\ss{}mann}, \citenamefont {Peruani},\ and\ \citenamefont {B\"ar}}]{PhysRevE.94.050602}%
  \BibitemOpen
  \bibfield  {author} {\bibinfo {author} {\bibfnamefont {R.}~\bibnamefont {Gro\ss{}mann}}, \bibinfo {author} {\bibfnamefont {F.}~\bibnamefont {Peruani}},\ and\ \bibinfo {author} {\bibfnamefont {M.}~\bibnamefont {B\"ar}},\ }\bibfield  {title} {\bibinfo {title} {Mesoscale pattern formation of self-propelled rods with velocity reversal},\ }\href {https://doi.org/10.1103/PhysRevE.94.050602} {\bibfield  {journal} {\bibinfo  {journal} {Phys. Rev. E}\ }\textbf {\bibinfo {volume} {94}},\ \bibinfo {pages} {050602} (\bibinfo {year} {2016})}\BibitemShut {NoStop}%
\bibitem [{\citenamefont {Nishiguchi}\ \emph {et~al.}(2017)\citenamefont {Nishiguchi}, \citenamefont {Nagai}, \citenamefont {Chat\'e},\ and\ \citenamefont {Sano}}]{PhysRevE.95.020601}%
  \BibitemOpen
  \bibfield  {author} {\bibinfo {author} {\bibfnamefont {D.}~\bibnamefont {Nishiguchi}}, \bibinfo {author} {\bibfnamefont {K.~H.}\ \bibnamefont {Nagai}}, \bibinfo {author} {\bibfnamefont {H.}~\bibnamefont {Chat\'e}},\ and\ \bibinfo {author} {\bibfnamefont {M.}~\bibnamefont {Sano}},\ }\bibfield  {title} {\bibinfo {title} {Long-range nematic order and anomalous fluctuations in suspensions of swimming filamentous bacteria},\ }\href {https://doi.org/10.1103/PhysRevE.95.020601} {\bibfield  {journal} {\bibinfo  {journal} {Phys. Rev. E}\ }\textbf {\bibinfo {volume} {95}},\ \bibinfo {pages} {020601} (\bibinfo {year} {2017})}\BibitemShut {NoStop}%
\bibitem [{\citenamefont {Genkin}\ \emph {et~al.}(2017)\citenamefont {Genkin}, \citenamefont {Sokolov}, \citenamefont {Lavrentovich},\ and\ \citenamefont {Aranson}}]{PhysRevX.7.011029}%
  \BibitemOpen
  \bibfield  {author} {\bibinfo {author} {\bibfnamefont {M.~M.}\ \bibnamefont {Genkin}}, \bibinfo {author} {\bibfnamefont {A.}~\bibnamefont {Sokolov}}, \bibinfo {author} {\bibfnamefont {O.~D.}\ \bibnamefont {Lavrentovich}},\ and\ \bibinfo {author} {\bibfnamefont {I.~S.}\ \bibnamefont {Aranson}},\ }\bibfield  {title} {\bibinfo {title} {Topological defects in a living nematic ensnare swimming bacteria},\ }\href {https://doi.org/10.1103/PhysRevX.7.011029} {\bibfield  {journal} {\bibinfo  {journal} {Phys. Rev. X}\ }\textbf {\bibinfo {volume} {7}},\ \bibinfo {pages} {011029} (\bibinfo {year} {2017})}\BibitemShut {NoStop}%
\bibitem [{\citenamefont {Ndlec}\ \emph {et~al.}(1997)\citenamefont {Ndlec}, \citenamefont {Surrey}, \citenamefont {Maggs},\ and\ \citenamefont {Leibler}}]{Ndlec1997}%
  \BibitemOpen
  \bibfield  {author} {\bibinfo {author} {\bibfnamefont {F.~J.}\ \bibnamefont {Ndlec}}, \bibinfo {author} {\bibfnamefont {T.}~\bibnamefont {Surrey}}, \bibinfo {author} {\bibfnamefont {A.~C.}\ \bibnamefont {Maggs}},\ and\ \bibinfo {author} {\bibfnamefont {S.}~\bibnamefont {Leibler}},\ }\bibfield  {title} {\bibinfo {title} {Self-organization of microtubules and motors},\ }\href {https://doi.org/10.1038/38532} {\bibfield  {journal} {\bibinfo  {journal} {Nature}\ }\textbf {\bibinfo {volume} {389}},\ \bibinfo {pages} {305} (\bibinfo {year} {1997})}\BibitemShut {NoStop}%
\bibitem [{\citenamefont {Butt}\ \emph {et~al.}(2010)\citenamefont {Butt}, \citenamefont {Mufti}, \citenamefont {Humayun}, \citenamefont {Rosenthal}, \citenamefont {Khan}, \citenamefont {Khan},\ and\ \citenamefont {Molloy}}]{Butt2010}%
  \BibitemOpen
  \bibfield  {author} {\bibinfo {author} {\bibfnamefont {T.}~\bibnamefont {Butt}}, \bibinfo {author} {\bibfnamefont {T.}~\bibnamefont {Mufti}}, \bibinfo {author} {\bibfnamefont {A.}~\bibnamefont {Humayun}}, \bibinfo {author} {\bibfnamefont {P.~B.}\ \bibnamefont {Rosenthal}}, \bibinfo {author} {\bibfnamefont {S.}~\bibnamefont {Khan}}, \bibinfo {author} {\bibfnamefont {S.}~\bibnamefont {Khan}},\ and\ \bibinfo {author} {\bibfnamefont {J.~E.}\ \bibnamefont {Molloy}},\ }\bibfield  {title} {\bibinfo {title} {Myosin motors drive long range alignment of actin filaments},\ }\href {https://doi.org/10.1074/jbc.m109.044792} {\bibfield  {journal} {\bibinfo  {journal} {Journal of Biological Chemistry}\ }\textbf {\bibinfo {volume} {285}},\ \bibinfo {pages} {4964–4974} (\bibinfo {year} {2010})}\BibitemShut {NoStop}%
\bibitem [{\citenamefont {Creppy}\ \emph {et~al.}(2015)\citenamefont {Creppy}, \citenamefont {Praud}, \citenamefont {Druart}, \citenamefont {Kohnke},\ and\ \citenamefont {Plourabou\'e}}]{PhysRevE.92.032722}%
  \BibitemOpen
  \bibfield  {author} {\bibinfo {author} {\bibfnamefont {A.}~\bibnamefont {Creppy}}, \bibinfo {author} {\bibfnamefont {O.}~\bibnamefont {Praud}}, \bibinfo {author} {\bibfnamefont {X.}~\bibnamefont {Druart}}, \bibinfo {author} {\bibfnamefont {P.~L.}\ \bibnamefont {Kohnke}},\ and\ \bibinfo {author} {\bibfnamefont {F.}~\bibnamefont {Plourabou\'e}},\ }\bibfield  {title} {\bibinfo {title} {Turbulence of swarming sperm},\ }\href {https://doi.org/10.1103/PhysRevE.92.032722} {\bibfield  {journal} {\bibinfo  {journal} {Phys. Rev. E}\ }\textbf {\bibinfo {volume} {92}},\ \bibinfo {pages} {032722} (\bibinfo {year} {2015})}\BibitemShut {NoStop}%
\bibitem [{\citenamefont {Blanch-Mercader}\ \emph {et~al.}(2018)\citenamefont {Blanch-Mercader}, \citenamefont {Yashunsky}, \citenamefont {Garcia}, \citenamefont {Duclos}, \citenamefont {Giomi},\ and\ \citenamefont {Silberzan}}]{Epithelial1}%
  \BibitemOpen
  \bibfield  {author} {\bibinfo {author} {\bibfnamefont {C.}~\bibnamefont {Blanch-Mercader}}, \bibinfo {author} {\bibfnamefont {V.}~\bibnamefont {Yashunsky}}, \bibinfo {author} {\bibfnamefont {S.}~\bibnamefont {Garcia}}, \bibinfo {author} {\bibfnamefont {G.}~\bibnamefont {Duclos}}, \bibinfo {author} {\bibfnamefont {L.}~\bibnamefont {Giomi}},\ and\ \bibinfo {author} {\bibfnamefont {P.}~\bibnamefont {Silberzan}},\ }\bibfield  {title} {\bibinfo {title} {Turbulent dynamics of epithelial cell cultures},\ }\href {https://doi.org/10.1103/PhysRevLett.120.208101} {\bibfield  {journal} {\bibinfo  {journal} {Phys. Rev. Lett.}\ }\textbf {\bibinfo {volume} {120}},\ \bibinfo {pages} {208101} (\bibinfo {year} {2018})}\BibitemShut {NoStop}%
\bibitem [{\citenamefont {Poujade}\ \emph {et~al.}(2007)\citenamefont {Poujade}, \citenamefont {Grasland-Mongrain}, \citenamefont {Hertzog}, \citenamefont {Jouanneau}, \citenamefont {Chavrier}, \citenamefont {Ladoux}, \citenamefont {Buguin},\ and\ \citenamefont {Silberzan}}]{Epithelial2}%
  \BibitemOpen
  \bibfield  {author} {\bibinfo {author} {\bibfnamefont {M.}~\bibnamefont {Poujade}}, \bibinfo {author} {\bibfnamefont {E.}~\bibnamefont {Grasland-Mongrain}}, \bibinfo {author} {\bibfnamefont {A.}~\bibnamefont {Hertzog}}, \bibinfo {author} {\bibfnamefont {J.}~\bibnamefont {Jouanneau}}, \bibinfo {author} {\bibfnamefont {P.}~\bibnamefont {Chavrier}}, \bibinfo {author} {\bibfnamefont {B.}~\bibnamefont {Ladoux}}, \bibinfo {author} {\bibfnamefont {A.}~\bibnamefont {Buguin}},\ and\ \bibinfo {author} {\bibfnamefont {P.}~\bibnamefont {Silberzan}},\ }\bibfield  {title} {\bibinfo {title} {Collective migration of an epithelial monolayer in response to a model wound},\ }\href {https://doi.org/10.1073/pnas.0705062104} {\bibfield  {journal} {\bibinfo  {journal} {Proceedings of the National Academy of Sciences}\ }\textbf {\bibinfo {volume} {104}},\ \bibinfo {pages} {15988–15993} (\bibinfo {year} {2007})}\BibitemShut {NoStop}%
\bibitem [{\citenamefont {Saw}\ \emph {et~al.}(2017{\natexlab{b}})\citenamefont {Saw}, \citenamefont {Doostmohammadi}, \citenamefont {Nier}, \citenamefont {Kocgozlu}, \citenamefont {Thampi}, \citenamefont {Toyama}, \citenamefont {Marcq}, \citenamefont {Lim}, \citenamefont {Yeomans},\ and\ \citenamefont {Ladoux}}]{Epithelial3}%
  \BibitemOpen
  \bibfield  {author} {\bibinfo {author} {\bibfnamefont {T.~B.}\ \bibnamefont {Saw}}, \bibinfo {author} {\bibfnamefont {A.}~\bibnamefont {Doostmohammadi}}, \bibinfo {author} {\bibfnamefont {V.}~\bibnamefont {Nier}}, \bibinfo {author} {\bibfnamefont {L.}~\bibnamefont {Kocgozlu}}, \bibinfo {author} {\bibfnamefont {S.}~\bibnamefont {Thampi}}, \bibinfo {author} {\bibfnamefont {Y.}~\bibnamefont {Toyama}}, \bibinfo {author} {\bibfnamefont {P.}~\bibnamefont {Marcq}}, \bibinfo {author} {\bibfnamefont {C.~T.}\ \bibnamefont {Lim}}, \bibinfo {author} {\bibfnamefont {J.~M.}\ \bibnamefont {Yeomans}},\ and\ \bibinfo {author} {\bibfnamefont {B.}~\bibnamefont {Ladoux}},\ }\bibfield  {title} {\bibinfo {title} {Topological defects in epithelia govern cell death and extrusion},\ }\href {https://doi.org/10.1038/nature21718} {\bibfield  {journal} {\bibinfo  {journal} {Nature}\ }\textbf {\bibinfo {volume} {544}},\ \bibinfo {pages} {212} (\bibinfo {year} {2017}{\natexlab{b}})}\BibitemShut {NoStop}%
\bibitem [{\citenamefont {Beris}\ and\ \citenamefont {Edwards}(1994)}]{beris1994thermodynamics}%
  \BibitemOpen
  \bibfield  {author} {\bibinfo {author} {\bibfnamefont {A.}~\bibnamefont {Beris}}\ and\ \bibinfo {author} {\bibfnamefont {B.}~\bibnamefont {Edwards}},\ }\href {https://books.google.nl/books?id=dqxFUy7_vhsC} {\emph {\bibinfo {title} {Thermodynamics of Flowing Systems: with Internal Microstructure}}},\ Oxford Engineering Science Series\ (\bibinfo  {publisher} {Oxford University Press},\ \bibinfo {year} {1994})\BibitemShut {NoStop}%
\bibitem [{\citenamefont {Aditi~Simha}\ and\ \citenamefont {Ramaswamy}(2002)}]{PhysRevLett.89.058101}%
  \BibitemOpen
  \bibfield  {author} {\bibinfo {author} {\bibfnamefont {R.}~\bibnamefont {Aditi~Simha}}\ and\ \bibinfo {author} {\bibfnamefont {S.}~\bibnamefont {Ramaswamy}},\ }\bibfield  {title} {\bibinfo {title} {Hydrodynamic fluctuations and instabilities in ordered suspensions of self-propelled particles},\ }\href {https://doi.org/10.1103/PhysRevLett.89.058101} {\bibfield  {journal} {\bibinfo  {journal} {Phys. Rev. Lett.}\ }\textbf {\bibinfo {volume} {89}},\ \bibinfo {pages} {058101} (\bibinfo {year} {2002})}\BibitemShut {NoStop}%
\bibitem [{\citenamefont {Kruse}\ \emph {et~al.}(2004)\citenamefont {Kruse}, \citenamefont {Joanny}, \citenamefont {J\"ulicher}, \citenamefont {Prost},\ and\ \citenamefont {Sekimoto}}]{PhysRevLett.92.078101}%
  \BibitemOpen
  \bibfield  {author} {\bibinfo {author} {\bibfnamefont {K.}~\bibnamefont {Kruse}}, \bibinfo {author} {\bibfnamefont {J.~F.}\ \bibnamefont {Joanny}}, \bibinfo {author} {\bibfnamefont {F.}~\bibnamefont {J\"ulicher}}, \bibinfo {author} {\bibfnamefont {J.}~\bibnamefont {Prost}},\ and\ \bibinfo {author} {\bibfnamefont {K.}~\bibnamefont {Sekimoto}},\ }\bibfield  {title} {\bibinfo {title} {Asters, vortices, and rotating spirals in active gels of polar filaments},\ }\href {https://doi.org/10.1103/PhysRevLett.92.078101} {\bibfield  {journal} {\bibinfo  {journal} {Phys. Rev. Lett.}\ }\textbf {\bibinfo {volume} {92}},\ \bibinfo {pages} {078101} (\bibinfo {year} {2004})}\BibitemShut {NoStop}%
\bibitem [{\citenamefont {Pedley}\ and\ \citenamefont {Kessler}(1992)}]{annurev:/content/journals/10.1146/annurev.fl.24.010192.001525}%
  \BibitemOpen
  \bibfield  {author} {\bibinfo {author} {\bibfnamefont {T.~J.}\ \bibnamefont {Pedley}}\ and\ \bibinfo {author} {\bibfnamefont {J.~O.}\ \bibnamefont {Kessler}},\ }\bibfield  {title} {\bibinfo {title} {Hydrodynamic phenomena in suspensions of swimming microorganisms},\ }\href {https://doi.org/https://doi.org/10.1146/annurev.fl.24.010192.001525} {\bibfield  {journal} {\bibinfo  {journal} {Annual Review of Fluid Mechanics}\ }\textbf {\bibinfo {volume} {24}},\ \bibinfo {pages} {313} (\bibinfo {year} {1992})}\BibitemShut {NoStop}%
\bibitem [{\citenamefont {{Martin}}\ \emph {et~al.}(1972)\citenamefont {{Martin}}, \citenamefont {{Parodi}},\ and\ \citenamefont {{Pershan}}}]{1972PhRvA...6.2401M}%
  \BibitemOpen
  \bibfield  {author} {\bibinfo {author} {\bibfnamefont {P.~C.}\ \bibnamefont {{Martin}}}, \bibinfo {author} {\bibfnamefont {O.}~\bibnamefont {{Parodi}}},\ and\ \bibinfo {author} {\bibfnamefont {P.~S.}\ \bibnamefont {{Pershan}}},\ }\bibfield  {title} {\bibinfo {title} {{Unified Hydrodynamic Theory for Crystals, Liquid Crystals, and Normal Fluids}},\ }\href {https://doi.org/10.1103/PhysRevA.6.2401} {\bibfield  {journal} {\bibinfo  {journal} {\pra}\ }\textbf {\bibinfo {volume} {6}},\ \bibinfo {pages} {2401} (\bibinfo {year} {1972})}\BibitemShut {NoStop}%
\bibitem [{\citenamefont {De~Groot}\ and\ \citenamefont {Mazur}(2013)}]{grootmazur}%
  \BibitemOpen
  \bibfield  {author} {\bibinfo {author} {\bibfnamefont {S.}~\bibnamefont {De~Groot}}\ and\ \bibinfo {author} {\bibfnamefont {P.}~\bibnamefont {Mazur}},\ }\href {https://books.google.nl/books?id=mfFyG9jfaMYC} {\emph {\bibinfo {title} {Non-Equilibrium Thermodynamics}}},\ Dover Books on Physics\ (\bibinfo  {publisher} {Dover Publications},\ \bibinfo {year} {2013})\BibitemShut {NoStop}%
\bibitem [{\citenamefont {Julicher}\ \emph {et~al.}(2018)\citenamefont {Julicher}, \citenamefont {Grill},\ and\ \citenamefont {Salbreux}}]{J_licher_2018}%
  \BibitemOpen
  \bibfield  {author} {\bibinfo {author} {\bibfnamefont {F.}~\bibnamefont {Julicher}}, \bibinfo {author} {\bibfnamefont {S.~W.}\ \bibnamefont {Grill}},\ and\ \bibinfo {author} {\bibfnamefont {G.}~\bibnamefont {Salbreux}},\ }\bibfield  {title} {\bibinfo {title} {Hydrodynamic theory of active matter},\ }\href {https://doi.org/10.1088/1361-6633/aab6bb} {\bibfield  {journal} {\bibinfo  {journal} {Reports on Progress in Physics}\ }\textbf {\bibinfo {volume} {81}},\ \bibinfo {pages} {076601} (\bibinfo {year} {2018})}\BibitemShut {NoStop}%
\bibitem [{\citenamefont {Parmeggiani}\ \emph {et~al.}(1999)\citenamefont {Parmeggiani}, \citenamefont {J\"ulicher}, \citenamefont {Ajdari},\ and\ \citenamefont {Prost}}]{PhysRevE.60.2127}%
  \BibitemOpen
  \bibfield  {author} {\bibinfo {author} {\bibfnamefont {A.}~\bibnamefont {Parmeggiani}}, \bibinfo {author} {\bibfnamefont {F.}~\bibnamefont {J\"ulicher}}, \bibinfo {author} {\bibfnamefont {A.}~\bibnamefont {Ajdari}},\ and\ \bibinfo {author} {\bibfnamefont {J.}~\bibnamefont {Prost}},\ }\bibfield  {title} {\bibinfo {title} {Energy transduction of isothermal ratchets: Generic aspects and specific examples close to and far from equilibrium},\ }\href {https://doi.org/10.1103/PhysRevE.60.2127} {\bibfield  {journal} {\bibinfo  {journal} {Phys. Rev. E}\ }\textbf {\bibinfo {volume} {60}},\ \bibinfo {pages} {2127} (\bibinfo {year} {1999})}\BibitemShut {NoStop}%
\bibitem [{\citenamefont {Armas}\ \emph {et~al.}(2024)\citenamefont {Armas}, \citenamefont {Jain},\ and\ \citenamefont {Lier}}]{Armas:2024iuy}%
  \BibitemOpen
  \bibfield  {author} {\bibinfo {author} {\bibfnamefont {J.}~\bibnamefont {Armas}}, \bibinfo {author} {\bibfnamefont {A.}~\bibnamefont {Jain}},\ and\ \bibinfo {author} {\bibfnamefont {R.}~\bibnamefont {Lier}},\ }\bibfield  {title} {\bibinfo {title} {{Hydrodynamics of thermal active matter}},\ }\href@noop {} {\bibfield  {journal} {\bibinfo  {journal} {preprint}\ } (\bibinfo {year} {2024})},\ \Eprint {https://arxiv.org/abs/2405.11023} {arXiv:2405.11023 [cond-mat.soft]} \BibitemShut {NoStop}%
\bibitem [{\citenamefont {Jülicher}\ \emph {et~al.}(2007)\citenamefont {Jülicher}, \citenamefont {Kruse}, \citenamefont {Prost},\ and\ \citenamefont {Joanny}}]{JULICHER20073}%
  \BibitemOpen
  \bibfield  {author} {\bibinfo {author} {\bibfnamefont {F.}~\bibnamefont {Jülicher}}, \bibinfo {author} {\bibfnamefont {K.}~\bibnamefont {Kruse}}, \bibinfo {author} {\bibfnamefont {J.}~\bibnamefont {Prost}},\ and\ \bibinfo {author} {\bibfnamefont {J.-F.}\ \bibnamefont {Joanny}},\ }\bibfield  {title} {\bibinfo {title} {Active behavior of the cytoskeleton},\ }\href {https://doi.org/https://doi.org/10.1016/j.physrep.2007.02.018} {\bibfield  {journal} {\bibinfo  {journal} {Physics Reports}\ }\textbf {\bibinfo {volume} {449}},\ \bibinfo {pages} {3} (\bibinfo {year} {2007})},\ \bibinfo {note} {nonequilibrium physics: From complex fluids to biological systems III. Living systems}\BibitemShut {NoStop}%
\bibitem [{\citenamefont {Callan-Jones}\ and\ \citenamefont {Jülicher}(2011)}]{Callan-Jones_2011}%
  \BibitemOpen
  \bibfield  {author} {\bibinfo {author} {\bibfnamefont {A.~C.}\ \bibnamefont {Callan-Jones}}\ and\ \bibinfo {author} {\bibfnamefont {F.}~\bibnamefont {Jülicher}},\ }\bibfield  {title} {\bibinfo {title} {Hydrodynamics of active permeating gels},\ }\href {https://doi.org/10.1088/1367-2630/13/9/093027} {\bibfield  {journal} {\bibinfo  {journal} {New Journal of Physics}\ }\textbf {\bibinfo {volume} {13}},\ \bibinfo {pages} {093027} (\bibinfo {year} {2011})}\BibitemShut {NoStop}%
\bibitem [{\citenamefont {Basu}\ \emph {et~al.}(2008)\citenamefont {Basu}, \citenamefont {Joanny}, \citenamefont {J{\"u}licher},\ and\ \citenamefont {Prost}}]{Basu2008}%
  \BibitemOpen
  \bibfield  {author} {\bibinfo {author} {\bibfnamefont {A.}~\bibnamefont {Basu}}, \bibinfo {author} {\bibfnamefont {J.~F.}\ \bibnamefont {Joanny}}, \bibinfo {author} {\bibfnamefont {F.}~\bibnamefont {J{\"u}licher}},\ and\ \bibinfo {author} {\bibfnamefont {J.}~\bibnamefont {Prost}},\ }\bibfield  {title} {\bibinfo {title} {Thermal and non-thermal fluctuations in active polar gels},\ }\href {https://doi.org/10.1140/epje/i2008-10364-9} {\bibfield  {journal} {\bibinfo  {journal} {The European Physical Journal E}\ }\textbf {\bibinfo {volume} {27}},\ \bibinfo {pages} {149} (\bibinfo {year} {2008})}\BibitemShut {NoStop}%
\bibitem [{\citenamefont {Ramaswamy}\ \emph {et~al.}(2003)\citenamefont {Ramaswamy}, \citenamefont {Simha},\ and\ \citenamefont {Toner}}]{SRamaswamy_2003}%
  \BibitemOpen
  \bibfield  {author} {\bibinfo {author} {\bibfnamefont {S.}~\bibnamefont {Ramaswamy}}, \bibinfo {author} {\bibfnamefont {R.~A.}\ \bibnamefont {Simha}},\ and\ \bibinfo {author} {\bibfnamefont {J.}~\bibnamefont {Toner}},\ }\bibfield  {title} {\bibinfo {title} {Active nematics on a substrate: Giant number fluctuations and long-time tails},\ }\href {https://doi.org/10.1209/epl/i2003-00346-7} {\bibfield  {journal} {\bibinfo  {journal} {Europhysics Letters}\ }\textbf {\bibinfo {volume} {62}},\ \bibinfo {pages} {196} (\bibinfo {year} {2003})}\BibitemShut {NoStop}%
\bibitem [{\citenamefont {Shankar}\ \emph {et~al.}(2018)\citenamefont {Shankar}, \citenamefont {Ramaswamy},\ and\ \citenamefont {Marchetti}}]{PhysRevE.97.012707}%
  \BibitemOpen
  \bibfield  {author} {\bibinfo {author} {\bibfnamefont {S.}~\bibnamefont {Shankar}}, \bibinfo {author} {\bibfnamefont {S.}~\bibnamefont {Ramaswamy}},\ and\ \bibinfo {author} {\bibfnamefont {M.~C.}\ \bibnamefont {Marchetti}},\ }\bibfield  {title} {\bibinfo {title} {Low-noise phase of a two-dimensional active nematic system},\ }\href {https://doi.org/10.1103/PhysRevE.97.012707} {\bibfield  {journal} {\bibinfo  {journal} {Phys. Rev. E}\ }\textbf {\bibinfo {volume} {97}},\ \bibinfo {pages} {012707} (\bibinfo {year} {2018})}\BibitemShut {NoStop}%
\bibitem [{\citenamefont {Voituriez}\ \emph {et~al.}(2005)\citenamefont {Voituriez}, \citenamefont {Joanny},\ and\ \citenamefont {Prost}}]{Voituriez_2005}%
  \BibitemOpen
  \bibfield  {author} {\bibinfo {author} {\bibfnamefont {R.}~\bibnamefont {Voituriez}}, \bibinfo {author} {\bibfnamefont {J.~F.}\ \bibnamefont {Joanny}},\ and\ \bibinfo {author} {\bibfnamefont {J.}~\bibnamefont {Prost}},\ }\bibfield  {title} {\bibinfo {title} {Spontaneous flow transition in active polar gels},\ }\href {https://doi.org/10.1209/epl/i2004-10501-2} {\bibfield  {journal} {\bibinfo  {journal} {Europhysics Letters}\ }\textbf {\bibinfo {volume} {70}},\ \bibinfo {pages} {404} (\bibinfo {year} {2005})}\BibitemShut {NoStop}%
\bibitem [{Note1()}]{Note1}%
  \BibitemOpen
  \bibinfo {note} {A work that appeared recently computes the thermal profile for pipe flows in the presence of an imposed activity gradient which manifests as a pressure gradient \cite {10.1063/5.0258996}. The key difference with our approach to temperature is that the energy balance equation we consider accounts for the fuel consumption that gives rise to activity. Additionally, we consider a different type of active matter and assume that it is only activity that drives the system out of equilibrium.}\BibitemShut {Stop}%
\bibitem [{\citenamefont {Landau}\ and\ \citenamefont {Lifshitz}(1959)}]{landau1959fluid}%
  \BibitemOpen
  \bibfield  {author} {\bibinfo {author} {\bibfnamefont {L.}~\bibnamefont {Landau}}\ and\ \bibinfo {author} {\bibfnamefont {E.}~\bibnamefont {Lifshitz}},\ }\href {https://books.google.co.uk/books?id=CVbntgAACAAJ} {\emph {\bibinfo {title} {{Fluid Mechanics}}}},\ Teoreticheskaia fizika\ (\bibinfo  {publisher} {Pergamon Press},\ \bibinfo {year} {1959})\BibitemShut {NoStop}%
\bibitem [{\citenamefont {Landau}\ and\ \citenamefont {Lifshitz}(1987)}]{1987xi}%
  \BibitemOpen
  \bibfield  {author} {\bibinfo {author} {\bibfnamefont {L.}~\bibnamefont {Landau}}\ and\ \bibinfo {author} {\bibfnamefont {E.}~\bibnamefont {Lifshitz}},\ }\href@noop {} {\emph {\bibinfo {title} {Fluid Mechanics (Second Edition)}}},\ \bibinfo {edition} {second edition}\ ed.\ (\bibinfo  {publisher} {Pergamon},\ \bibinfo {year} {1987})\BibitemShut {NoStop}%
\bibitem [{\citenamefont {Fox}\ and\ \citenamefont {Uhlenbeck}(1970)}]{Fox1970}%
  \BibitemOpen
  \bibfield  {author} {\bibinfo {author} {\bibfnamefont {R.~F.}\ \bibnamefont {Fox}}\ and\ \bibinfo {author} {\bibfnamefont {G.~E.}\ \bibnamefont {Uhlenbeck}},\ }\bibfield  {title} {\bibinfo {title} {Contributions to non-equilibrium thermodynamics. i. theory of hydrodynamical fluctuations},\ }\href {https://doi.org/10.1063/1.1693183} {\bibfield  {journal} {\bibinfo  {journal} {The Physics of Fluids}\ }\textbf {\bibinfo {volume} {13}},\ \bibinfo {pages} {1893–1902} (\bibinfo {year} {1970})}\BibitemShut {NoStop}%
\bibitem [{\citenamefont {Machlup}\ and\ \citenamefont {Onsager}(1953)}]{PhysRev.91.1512}%
  \BibitemOpen
  \bibfield  {author} {\bibinfo {author} {\bibfnamefont {S.}~\bibnamefont {Machlup}}\ and\ \bibinfo {author} {\bibfnamefont {L.}~\bibnamefont {Onsager}},\ }\bibfield  {title} {\bibinfo {title} {Fluctuations and irreversible process. ii. systems with kinetic energy},\ }\href {https://doi.org/10.1103/PhysRev.91.1512} {\bibfield  {journal} {\bibinfo  {journal} {Phys. Rev.}\ }\textbf {\bibinfo {volume} {91}},\ \bibinfo {pages} {1512} (\bibinfo {year} {1953})}\BibitemShut {NoStop}%
\bibitem [{\citenamefont {Chandrasekhar}(1981)}]{chandrasekhar1981hydrodynamic}%
  \BibitemOpen
  \bibfield  {author} {\bibinfo {author} {\bibfnamefont {S.}~\bibnamefont {Chandrasekhar}},\ }\href {https://books.google.nl/books?id=oU_-6ikmidoC} {\emph {\bibinfo {title} {Hydrodynamic and Hydromagnetic Stability}}},\ Dover Books on Physics Series\ (\bibinfo  {publisher} {Dover Publications},\ \bibinfo {year} {1981})\BibitemShut {NoStop}%
\bibitem [{Note2()}]{Note2}%
  \BibitemOpen
  \bibinfo {note} {The energy balance equation can also be recast in terms of total energy, including the fluid and fuel contributions, which is conserved when the outflow of energy to the environment is absent~\cite {J_licher_2018}. However, the non-conservative coupling to the environment is crucial to maintain active steady states; see appendix A of~\cite {Armas:2024iuy} for more details.}\BibitemShut {Stop}%
\bibitem [{\citenamefont {de~Zarate}\ and\ \citenamefont {Sengers}(2006)}]{sengersortiz}%
  \BibitemOpen
  \bibfield  {author} {\bibinfo {author} {\bibfnamefont {J.}~\bibnamefont {de~Zarate}}\ and\ \bibinfo {author} {\bibfnamefont {J.}~\bibnamefont {Sengers}},\ }\href {https://books.google.nl/books?id=4sx3CVMreJMC} {\emph {\bibinfo {title} {Hydrodynamic Fluctuations in Fluids and Fluid Mixtures}}}\ (\bibinfo  {publisher} {Elsevier Science},\ \bibinfo {year} {2006})\BibitemShut {NoStop}%
\bibitem [{\citenamefont {Olmsted}\ and\ \citenamefont {Goldbart}(1992)}]{PhysRevA.46.4966}%
  \BibitemOpen
  \bibfield  {author} {\bibinfo {author} {\bibfnamefont {P.~D.}\ \bibnamefont {Olmsted}}\ and\ \bibinfo {author} {\bibfnamefont {P.~M.}\ \bibnamefont {Goldbart}},\ }\bibfield  {title} {\bibinfo {title} {Isotropic-nematic transition in shear flow: State selection, coexistence, phase transitions, and critical behavior},\ }\href {https://doi.org/10.1103/PhysRevA.46.4966} {\bibfield  {journal} {\bibinfo  {journal} {Phys. Rev. A}\ }\textbf {\bibinfo {volume} {46}},\ \bibinfo {pages} {4966} (\bibinfo {year} {1992})}\BibitemShut {NoStop}%
\bibitem [{\citenamefont {Giomi}\ \emph {et~al.}(2012)\citenamefont {Giomi}, \citenamefont {Mahadevan}, \citenamefont {Chakraborty},\ and\ \citenamefont {Hagan}}]{Giomi_2012}%
  \BibitemOpen
  \bibfield  {author} {\bibinfo {author} {\bibfnamefont {L.}~\bibnamefont {Giomi}}, \bibinfo {author} {\bibfnamefont {L.}~\bibnamefont {Mahadevan}}, \bibinfo {author} {\bibfnamefont {B.}~\bibnamefont {Chakraborty}},\ and\ \bibinfo {author} {\bibfnamefont {M.~F.}\ \bibnamefont {Hagan}},\ }\bibfield  {title} {\bibinfo {title} {Banding, excitability and chaos in active nematic suspensions},\ }\href {https://doi.org/10.1088/0951-7715/25/8/2245} {\bibfield  {journal} {\bibinfo  {journal} {Nonlinearity}\ }\textbf {\bibinfo {volume} {25}},\ \bibinfo {pages} {2245} (\bibinfo {year} {2012})}\BibitemShut {NoStop}%
\bibitem [{Note3()}]{Note3}%
  \BibitemOpen
  \bibinfo {note} {In theory, one can also account for the active contribution $ \zeta $ in an Onsager reciprocal way by adding a mechanosensitive contribution to $r_{{\protect \mathsmaller {\protect \mathsf E}}}$. This possibility was considered in \cite {Armas:2024iuy}.}\BibitemShut {Stop}%
\bibitem [{\citenamefont {Thampi}\ \emph {et~al.}(2014)\citenamefont {Thampi}, \citenamefont {Golestanian},\ and\ \citenamefont {Yeomans}}]{PhysRevE.90.062307}%
  \BibitemOpen
  \bibfield  {author} {\bibinfo {author} {\bibfnamefont {S.~P.}\ \bibnamefont {Thampi}}, \bibinfo {author} {\bibfnamefont {R.}~\bibnamefont {Golestanian}},\ and\ \bibinfo {author} {\bibfnamefont {J.~M.}\ \bibnamefont {Yeomans}},\ }\bibfield  {title} {\bibinfo {title} {Active nematic materials with substrate friction},\ }\href {https://doi.org/10.1103/PhysRevE.90.062307} {\bibfield  {journal} {\bibinfo  {journal} {Phys. Rev. E}\ }\textbf {\bibinfo {volume} {90}},\ \bibinfo {pages} {062307} (\bibinfo {year} {2014})}\BibitemShut {NoStop}%
\bibitem [{\citenamefont {Giomi}\ \emph {et~al.}(2014)\citenamefont {Giomi}, \citenamefont {Bowick}, \citenamefont {Mishra}, \citenamefont {Sknepnek},\ and\ \citenamefont {Cristina~Marchetti}}]{giomiroyal}%
  \BibitemOpen
  \bibfield  {author} {\bibinfo {author} {\bibfnamefont {L.}~\bibnamefont {Giomi}}, \bibinfo {author} {\bibfnamefont {M.~J.}\ \bibnamefont {Bowick}}, \bibinfo {author} {\bibfnamefont {P.}~\bibnamefont {Mishra}}, \bibinfo {author} {\bibfnamefont {R.}~\bibnamefont {Sknepnek}},\ and\ \bibinfo {author} {\bibfnamefont {M.}~\bibnamefont {Cristina~Marchetti}},\ }\bibfield  {title} {\bibinfo {title} {Defect dynamics in active nematics},\ }\href {https://doi.org/10.1098/rsta.2013.0365} {\bibfield  {journal} {\bibinfo  {journal} {Philosophical Transactions of the Royal Society A: Mathematical, Physical and Engineering Sciences}\ }\textbf {\bibinfo {volume} {372}},\ \bibinfo {pages} {20130365} (\bibinfo {year} {2014})}\BibitemShut {NoStop}%
\bibitem [{\citenamefont {Ramaswamy}\ and\ \citenamefont {Rao}(2007)}]{Ramaswamy_2007}%
  \BibitemOpen
  \bibfield  {author} {\bibinfo {author} {\bibfnamefont {S.}~\bibnamefont {Ramaswamy}}\ and\ \bibinfo {author} {\bibfnamefont {M.}~\bibnamefont {Rao}},\ }\bibfield  {title} {\bibinfo {title} {Active-filament hydrodynamics: instabilities, boundary conditions and rheology},\ }\href {https://doi.org/10.1088/1367-2630/9/11/423} {\bibfield  {journal} {\bibinfo  {journal} {New Journal of Physics}\ }\textbf {\bibinfo {volume} {9}},\ \bibinfo {pages} {423} (\bibinfo {year} {2007})}\BibitemShut {NoStop}%
\bibitem [{\citenamefont {Li}\ \emph {et~al.}(1994)\citenamefont {Li}, \citenamefont {Segrè}, \citenamefont {Gammon},\ and\ \citenamefont {Sengers}}]{LI1994399}%
  \BibitemOpen
  \bibfield  {author} {\bibinfo {author} {\bibfnamefont {W.}~\bibnamefont {Li}}, \bibinfo {author} {\bibfnamefont {P.}~\bibnamefont {Segrè}}, \bibinfo {author} {\bibfnamefont {R.}~\bibnamefont {Gammon}},\ and\ \bibinfo {author} {\bibfnamefont {J.}~\bibnamefont {Sengers}},\ }\bibfield  {title} {\bibinfo {title} {Small-angle rayleigh scattering from nonequilibrium fluctuations in liquids and liquid mixtures},\ }\href {https://doi.org/https://doi.org/10.1016/0378-4371(94)90440-5} {\bibfield  {journal} {\bibinfo  {journal} {Physica A: Statistical Mechanics and its Applications}\ }\textbf {\bibinfo {volume} {204}},\ \bibinfo {pages} {399} (\bibinfo {year} {1994})}\BibitemShut {NoStop}%
\bibitem [{\citenamefont {{Lavi}}\ \emph {et~al.}(2024)\citenamefont {{Lavi}}, \citenamefont {{Alert}}, \citenamefont {{Joanny}},\ and\ \citenamefont {{Casademunt}}}]{2024arXiv240316841L}%
  \BibitemOpen
  \bibfield  {author} {\bibinfo {author} {\bibfnamefont {I.}~\bibnamefont {{Lavi}}}, \bibinfo {author} {\bibfnamefont {R.}~\bibnamefont {{Alert}}}, \bibinfo {author} {\bibfnamefont {J.-F.}\ \bibnamefont {{Joanny}}},\ and\ \bibinfo {author} {\bibfnamefont {J.}~\bibnamefont {{Casademunt}}},\ }\bibfield  {title} {\bibinfo {title} {{Nonlinear spontaneous flow instability in active nematics}},\ }\href {https://doi.org/10.48550/arXiv.2403.16841} {\bibfield  {journal} {\bibinfo  {journal} {arXiv e-prints}\ ,\ \bibinfo {eid} {arXiv:2403.16841}} (\bibinfo {year} {2024})},\ \Eprint {https://arxiv.org/abs/2403.16841} {arXiv:2403.16841 [cond-mat.soft]} \BibitemShut {NoStop}%
\bibitem [{Note4()}]{Note4}%
  \BibitemOpen
  \bibinfo {note} {This is assuming that the transition is continuous. For certain values of $\lambda $, the transition can become subcritical and discontinuous \cite {lavi2024nonlinearspontaneousflowinstability}.}\BibitemShut {Stop}%
\bibitem [{\citenamefont {Edwards}\ and\ \citenamefont {Yeomans}(2009)}]{Edwards_2009}%
  \BibitemOpen
  \bibfield  {author} {\bibinfo {author} {\bibfnamefont {S.~A.}\ \bibnamefont {Edwards}}\ and\ \bibinfo {author} {\bibfnamefont {J.~M.}\ \bibnamefont {Yeomans}},\ }\bibfield  {title} {\bibinfo {title} {Spontaneous flow states in active nematics: A unified picture},\ }\href {https://doi.org/10.1209/0295-5075/85/18008} {\bibfield  {journal} {\bibinfo  {journal} {Europhysics Letters}\ }\textbf {\bibinfo {volume} {85}},\ \bibinfo {pages} {18008} (\bibinfo {year} {2009})}\BibitemShut {NoStop}%
\bibitem [{\citenamefont {Marenduzzo}\ \emph {et~al.}(2007)\citenamefont {Marenduzzo}, \citenamefont {Orlandini}, \citenamefont {Cates},\ and\ \citenamefont {Yeomans}}]{PhysRevE.76.031921}%
  \BibitemOpen
  \bibfield  {author} {\bibinfo {author} {\bibfnamefont {D.}~\bibnamefont {Marenduzzo}}, \bibinfo {author} {\bibfnamefont {E.}~\bibnamefont {Orlandini}}, \bibinfo {author} {\bibfnamefont {M.~E.}\ \bibnamefont {Cates}},\ and\ \bibinfo {author} {\bibfnamefont {J.~M.}\ \bibnamefont {Yeomans}},\ }\bibfield  {title} {\bibinfo {title} {Steady-state hydrodynamic instabilities of active liquid crystals: Hybrid lattice boltzmann simulations},\ }\href {https://doi.org/10.1103/PhysRevE.76.031921} {\bibfield  {journal} {\bibinfo  {journal} {Phys. Rev. E}\ }\textbf {\bibinfo {volume} {76}},\ \bibinfo {pages} {031921} (\bibinfo {year} {2007})}\BibitemShut {NoStop}%
\bibitem [{\citenamefont {Duclos}\ \emph {et~al.}(2018)\citenamefont {Duclos}, \citenamefont {Blanch-Mercader}, \citenamefont {Yashunsky}, \citenamefont {Salbreux}, \citenamefont {Joanny}, \citenamefont {Prost},\ and\ \citenamefont {Silberzan}}]{duclos2018spontaneous}%
  \BibitemOpen
  \bibfield  {author} {\bibinfo {author} {\bibfnamefont {G.}~\bibnamefont {Duclos}}, \bibinfo {author} {\bibfnamefont {C.}~\bibnamefont {Blanch-Mercader}}, \bibinfo {author} {\bibfnamefont {V.}~\bibnamefont {Yashunsky}}, \bibinfo {author} {\bibfnamefont {G.}~\bibnamefont {Salbreux}}, \bibinfo {author} {\bibfnamefont {J.-F.}\ \bibnamefont {Joanny}}, \bibinfo {author} {\bibfnamefont {J.}~\bibnamefont {Prost}},\ and\ \bibinfo {author} {\bibfnamefont {P.}~\bibnamefont {Silberzan}},\ }\bibfield  {title} {\bibinfo {title} {Spontaneous shear flow in confined cellular nematics},\ }\href@noop {} {\bibfield  {journal} {\bibinfo  {journal} {Nature physics}\ }\textbf {\bibinfo {volume} {14}},\ \bibinfo {pages} {728} (\bibinfo {year} {2018})}\BibitemShut {NoStop}%
\bibitem [{\citenamefont {Chandrakar}\ \emph {et~al.}(2020)\citenamefont {Chandrakar}, \citenamefont {Varghese}, \citenamefont {Aghvami}, \citenamefont {Baskaran}, \citenamefont {Dogic},\ and\ \citenamefont {Duclos}}]{duclos2}%
  \BibitemOpen
  \bibfield  {author} {\bibinfo {author} {\bibfnamefont {P.}~\bibnamefont {Chandrakar}}, \bibinfo {author} {\bibfnamefont {M.}~\bibnamefont {Varghese}}, \bibinfo {author} {\bibfnamefont {S.}~\bibnamefont {Aghvami}}, \bibinfo {author} {\bibfnamefont {A.}~\bibnamefont {Baskaran}}, \bibinfo {author} {\bibfnamefont {Z.}~\bibnamefont {Dogic}},\ and\ \bibinfo {author} {\bibfnamefont {G.}~\bibnamefont {Duclos}},\ }\bibfield  {title} {\bibinfo {title} {Confinement controls the bend instability of three-dimensional active liquid crystals},\ }\href {https://doi.org/10.1103/PhysRevLett.125.257801} {\bibfield  {journal} {\bibinfo  {journal} {Phys. Rev. Lett.}\ }\textbf {\bibinfo {volume} {125}},\ \bibinfo {pages} {257801} (\bibinfo {year} {2020})}\BibitemShut {NoStop}%
\bibitem [{\citenamefont {Das}(2025)}]{10.1063/5.0258996}%
  \BibitemOpen
  \bibfield  {author} {\bibinfo {author} {\bibfnamefont {S.}~\bibnamefont {Das}},\ }\bibfield  {title} {\bibinfo {title} {Thermally fully developed pipe flows of active liquids},\ }\href {https://doi.org/10.1063/5.0258996} {\bibfield  {journal} {\bibinfo  {journal} {Physics of Fluids}\ }\textbf {\bibinfo {volume} {37}},\ \bibinfo {pages} {032110} (\bibinfo {year} {2025})},\ \Eprint {https://arxiv.org/abs/https://pubs.aip.org/aip/pof/article-pdf/doi/10.1063/5.0258996/20425931/032110\_1\_5.0258996.pdf} {https://pubs.aip.org/aip/pof/article-pdf/doi/10.1063/5.0258996/20425931/032110\_1\_5.0258996.pdf} \BibitemShut {NoStop}%
\bibitem [{\citenamefont {Lavi}\ \emph {et~al.}(2024)\citenamefont {Lavi}, \citenamefont {Alert}, \citenamefont {Joanny},\ and\ \citenamefont {Casademunt}}]{lavi2024nonlinearspontaneousflowinstability}%
  \BibitemOpen
  \bibfield  {author} {\bibinfo {author} {\bibfnamefont {I.}~\bibnamefont {Lavi}}, \bibinfo {author} {\bibfnamefont {R.}~\bibnamefont {Alert}}, \bibinfo {author} {\bibfnamefont {J.-F.}\ \bibnamefont {Joanny}},\ and\ \bibinfo {author} {\bibfnamefont {J.}~\bibnamefont {Casademunt}},\ }\href {https://arxiv.org/abs/2403.16841} {\bibinfo {title} {Nonlinear spontaneous flow instability in active nematics}} (\bibinfo {year} {2024}),\ \Eprint {https://arxiv.org/abs/2403.16841} {arXiv:2403.16841 [cond-mat.soft]} \BibitemShut {NoStop}%
\bibitem [{\citenamefont {Forster}(2018)}]{Forster2018}%
  \BibitemOpen
  \bibfield  {author} {\bibinfo {author} {\bibfnamefont {D.}~\bibnamefont {Forster}},\ }\href {https://doi.org/10.1201/9780429493683} {\emph {\bibinfo {title} {Hydrodynamic Fluctuations, Broken Symmetry, and Correlation Functions}}}\ (\bibinfo  {publisher} {CRC Press},\ \bibinfo {year} {2018})\BibitemShut {NoStop}%
\bibitem [{\citenamefont {Hatwalne}\ \emph {et~al.}(2004)\citenamefont {Hatwalne}, \citenamefont {Ramaswamy}, \citenamefont {Rao},\ and\ \citenamefont {Simha}}]{PhysRevLett.92.118101}%
  \BibitemOpen
  \bibfield  {author} {\bibinfo {author} {\bibfnamefont {Y.}~\bibnamefont {Hatwalne}}, \bibinfo {author} {\bibfnamefont {S.}~\bibnamefont {Ramaswamy}}, \bibinfo {author} {\bibfnamefont {M.}~\bibnamefont {Rao}},\ and\ \bibinfo {author} {\bibfnamefont {R.~A.}\ \bibnamefont {Simha}},\ }\bibfield  {title} {\bibinfo {title} {Rheology of active-particle suspensions},\ }\href {https://doi.org/10.1103/PhysRevLett.92.118101} {\bibfield  {journal} {\bibinfo  {journal} {Phys. Rev. Lett.}\ }\textbf {\bibinfo {volume} {92}},\ \bibinfo {pages} {118101} (\bibinfo {year} {2004})}\BibitemShut {NoStop}%
\end{thebibliography}
\end{document}